\documentclass[10pt, a4paper, oneside]{article}
\usepackage[hidelinks]{hyperref}
\usepackage{sci-paper}
\usepackage{times}
\usepackage{graphicx}
\usepackage{url}
\usepackage{ulem}
\usepackage{mathtools}
\usepackage{scalerel}
\usepackage{setspace}
\usepackage[strict]{changepage}
\usepackage{caption}
\usepackage[letterspace=-50]{microtype}
\usepackage{afterpage}
\usepackage{ragged2e}
\usepackage[ruled,vlined]{algorithm2e}
\usepackage{float}
\usepackage{hyperref}
\usepackage{titlesec}
\usepackage{comment}
 
\usepackage[margin=0cm, left=4.6cm, right=4.2cm, top=3.9cm, bottom=6.13cm, a4paper, headheight=0.5cm, headsep=0.5cm]{geometry}
\usepackage{fancyhdr}
\usepackage[format=plain, labelfont=it, textfont=it, justification=centering]{caption}
\usepackage{breakcites}
\usepackage{microtype}
\usepackage{geometry}
\usepackage{footmisc}
\usepackage{natbib}

\hypersetup{
    colorlinks=true,
    linkcolor=blue,
    filecolor=magenta,      
    urlcolor=cyan,
}

 \titleformat*{\section}{\Large\bfseries}
 \titleformat*{\subsection}{\normalsize\bfseries}
 \graphicspath{{./img/}}
 \apptocmd{\frame}{}{\justifying}{}
 \newcommand\startingPage{1}
 \newcommand\paperauthor{{University of Pisa. }}

 \newcommand\papertitle{Web image search engine based on\\ LSH index and CNN Resnet50}

 \header{\paperauthor \papertitle}

\begin{document}

 \title{{\fontsize{14pt}{14pt}\selectfont{\vspace*{-3mm}\papertitle\vspace*{-1mm}}}}

 \author{{\bfseries\fontsize{10pt}{10pt}\selectfont{Alice Nannini*, Marco Parola*, Stefano Poleggi \footnote[1]{Equal contribution. Listing order is random, all members cooperated providing important achievements.}}}\\
   {\fontsize{9pt}{12pt}\selectfont{University of Pisa, Department of Information Engineering, largo L. Lazzarino 1, 56122, Pisa, Italy
 }}
}

\maketitle

{\fontfamily{ptm}\selectfont
\begin{abstract}
{\fontsize{9pt}{9pt}\selectfont{\vspace*{-2mm}
To implement a good Content Based Image Retrieval (CBIR) system, it is essential to adopt efficient search methods. One way to achieve this results is by exploiting approximate search techniques. In fact, when we deal with very large collections of data, using an exact search method makes the system very slow.  In this project, we adopt the Locality Sensitive Hashing (LSH) index to implement a CBIR system that allows us to perform fast similarity search on deep features. Specifically, we exploit transfer learning techniques to extract deep features from images; this phase is done using two famous Convolutional Neural Networks (CNNs) as features extractors: Resnet50 and Resnet50v2, both pre-trained on ImageNet. Then we try out several fully connected deep neural networks, built on top of both of the previously mentioned CNNs in order to fine-tuned them on our dataset. In both of previous cases, we index the features within our LSH index implementation and within a sequential scan, to better understand how much the introduction of the index affects the results. Finally, we carry out a performance analysis: we evaluate the relevance of the result set, computing the mAP (mean Average Precision) value obtained during the different experiments with respect to the number of done comparison and varying the hyper-parameter values of the LSH index.}}
\end{abstract}}

{\fontfamily{ptm}\selectfont
\begin{keywords}
{\fontsize{9pt}{9pt}\selectfont{
Resnet50 -- Information Retrieval -- Similarity Search -- Classification -- Computer Vision -- Locality Similarity Hashing -- Convolutional Neural Network -- Image Analysis -- ImageNet}}
\end{keywords}}

\vspace{10mm}

\section{Introduction}\label{sec:1}

Our project consists in designing and implementing a Web Search Engine that can be interrogated by the user by uploading an image that will be used as query. The system shows to the user the 10 most similar images as result.
We want to implement an index that guarantee a good performance in terms of relevant objects retrieved and small number of computed distances. Since these two aspects are in conflict with each other, we try to find the best trade-off between these two values: ideally we would like to display images very similar to the query, making few comparisons between objects.\\
The study is composed by six sections. In Section \ref{sec:2}{} we describe the dataset and the data pre-processing phase. In Section \ref{sec:3}{} we describe the implementation of the LSH index. In Section \ref{sec:4}{}, we develope two LSH indexes: one on top of the features extracted using Resnet50 and the other on top of the features extracted using Resnet50v2; we measure the performances of the approximate search and we compare them with respect to the ones measured on the sequential scan. The metrics we take into account to compare the performances are the relevance, measured by the mAP value, and the number of distance computation calculated during the search. In Section \ref{sec:5}{} we repeat the experiments of the Section \ref{sec:4}{} adopting a fine-tuning approach, in order to increase the accuracy. In Section \ref{sec:6}{} we show how the hyper-parameter values of the LSH index influence the performances. Finally, in Section \ref{sec:7}{} we implement a Web Search Engine based on the best index built during the previous sections and we develop a GUI to allow the user to query the engine via browser. \\
We specify that the project has been realized during the Multimedia and Information Retrieval exam of the master course in Artificial Intelligence and Data Engineering at the University of Pisa. In this regard, for the realization we referred to \cite{Goodfellow-et-al-2016},  \cite{zezula2006similarity}, and \cite{cambridge2009online}.

\newpage

\section{Dataset Analysis and Preprocessing}\label{sec:2}

The system is based on two datasets which serve different purposes: ArtImages \url{https://www.kaggle.com/thedownhill/art-images-drawings-painting-sculpture-engraving} and MIRFlickr \cite{huiskes08} datasets. 
The first one is separated on training and validation sets; each of them contains 5 categories of images, corresponding to the 5 classes: \textbf{Drawings, Engraving, Iconography, Painting, Sculpture}.
The dataset contained some corrupted images, which we have detected and deleted with a python script. After this removal task, the training set is composed by a total of \textbf{7786} images, while the validation set counts a total of \textbf{856} images.
The second dataset consists of 25.000 images downloaded from the social photography site Flickr through its public API. We exploit this dataset as a distractor for our CBIR system.
The following figures show the distribution of the classes of the training set and validation set. We can observe very similar distributions between the classes of the training set and validation set.

\begin{figure}[H]
\centering
	\begin{minipage}[c]{.47\textwidth}
		\centering\setlength{\captionmargin}{0pt}%
		\includegraphics[width=1.\textwidth]{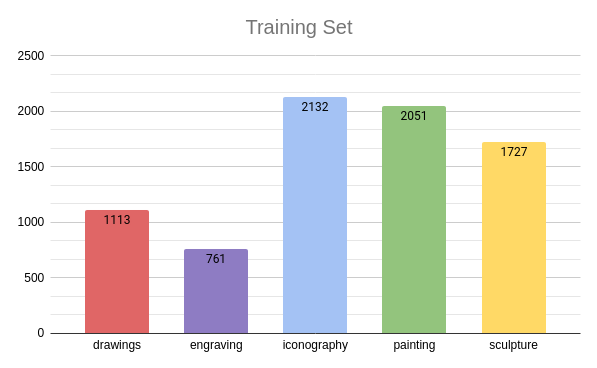}
		\label{fig:}
	\end{minipage}
	\hspace{5mm}%
	\begin{minipage}[c]{.47\textwidth}
		\centering\setlength{\captionmargin}{0pt}%
		\includegraphics[width=1.\textwidth]{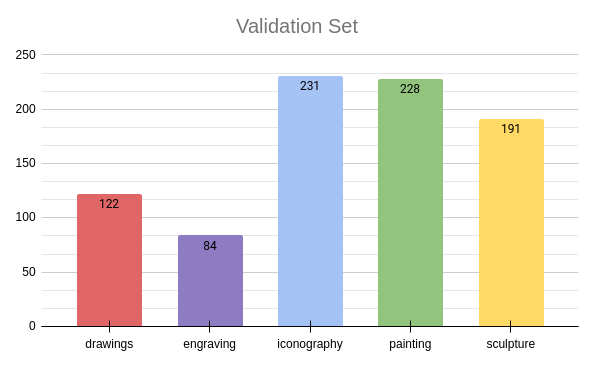}
		\label{fig:scratch_matrix}
	\end{minipage}%
	\caption{The dataset distribution of classes}
\end{figure}

\section{Locality Sensitive Hashing (LSH) Index}\label{sec:3}
At the core of our image search engine there is a \textbf{Locality Sensitive Hashing} (LSH) index \cite{datar2004locality} \cite{LSH}. We store data points into buckets after we hashed them, in such a way that data points near each other are located within the same bucket with high probability, while data points far from each other are likely to be in different buckets.
This could be done by building different hash functions \textbf{g()}, where each of them is defined exploiting the concept of random projections; more formally as
\begin{displaymath}
g(p) = < h_{1}(p),\ ... \ , h_{k}(p) >
\end{displaymath}
Where each element h\textsubscript{i}() is defined as
\begin{equation*}
h\textsubscript{i}(p) 
= \frac{(p * X\textsubscript{i} + b\textsubscript{i})}{w}  
\ \ \ \ \ \ \ i = 1 ... k
\end{equation*}
and represents a hyper-plane described by \textit{X\textsubscript{i}} and \textit{b}, while \textit{w} is the length of the segments in which the hyper-planes are divided.
We define \textit{L} hash functions, corresponding to \textit{L} buckets, then each image is inserted in a bucket for each of the \textit{L} hash functions.
When a query is passed to the search engine, its features are hashed using each \textit{g} function and the corresponding buckets are identified. The elements belonging to them are retrieved and sorted, in order to select the top-k more similar to the query.

\section{Features extraction}\label{sec:4}

Starting with \cite{sharif2014cnn}, state-of-the-art features for image retrieval have been extracted considering the activation of neurons in hidden layers of convolutional neural networks \cite{amato2016yfcc100m,amato2017searching,jogin2018feature}.

In this work, we extracted the features from the images using the pre-trained convolutional neural networks \textbf{Resnet50} and \textbf{Resnet50v2} trained on \textit{Imagenet} dataset, adopting a features extraction approach.\\
The strength of this convolutional neural network is the skip connection \cite{he2016deep}: the figure on the left is stacking convolution layers together one after the other. On the right we still stack convolution layers as before but we add also the original input to the output of the convolution block. This is called skip connection.\\
\begin{figure}[H]
    \centering
    \begin{minipage}{0.5\textwidth}
        \includegraphics[scale=0.46]{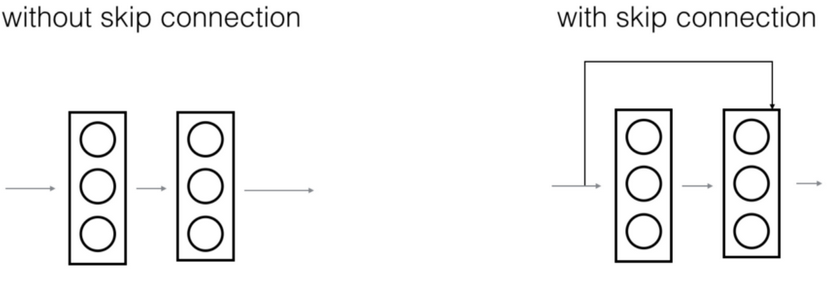}
    \end{minipage}
    \caption{Skip connection intuition}
\vspace{2mm}
\end{figure}

\subsection{Distance distribution}
We performed a quick analysis on the distribution of distances among the objects in our training-set (or rather among the features extracted from the objects). This was done so that we could better interpret the data that we will obtain in the following sections by applying the k-NN search algorithm. \\
First, we calculated the average distance between the features extracted from Resnet50 and Resnet50v2, both with the Euclidean distance and with the Cosine similarity metrics.

\begin{center}
\begin{tabular}{|l|cc|}
\hline
 &  mean Euclidean dist. & mean Cosine sim. \\
\hline
Resnet50   & 3.015  & 0.994  \\
Resnet50v2 & 31.467 & 0.3614 \\
\hline
\end{tabular}
\end{center}

\noindent To complete the analysis, we also calculated the average distance between features divided by class, in this case using only the Euclidean distance for brevity. The results are shown below.

\begin{center}
\begin{tabular}{|l|ccccc|}
\hline
mean Euclidean dist. & 0 - drawings & 1 - engraving & 2 - iconography & 3 - painting & 4 - sculpture \\
\hline
Resnet50   & 3.2472 & 4.3399  & 3.0041  & 2.5269 & 2.8741  \\
Resnet50v2 & 31.199 & 31.7557 & 30.0471 & 31.078 & 33.727  \\
\hline
\end{tabular}
\end{center}

\noindent Looking at both the tables, we can conclude that the features extracted with Resnet50 are much closer and similar than the features extracted with Resnet50v2. In addition, the distinct values by class are close to the mean value. This means that also in the search results we will get higher distance values for the second network with respect to the first one.

\subsection{Resnet50}
\subsubsection{Sequential search after Features Extraction, using Resnet50}
During this section we extracted the features from the dataset using the pre-trained network Resnet50, adopting a features extraction technique and we computed a sequential scan \textit{both with and without the distractor}, in order to compare the performance when we introduced some images as noise.\\
We obtained the following results:\\

Test without distractor
\begin{center}
\begin{tabular}{|l|cc|cc|c|}
\hline
k = 10 &  mAP euclidean & avg dist & mAP cosine & avg sim & $\#$Items\\
\hline
Resnet50 & 0.715 & 1.085 & 0.718 & 0.999 & 7784 \\
\hline
\end{tabular}
\end{center}

\begin{center}
\begin{tabular}{|l|ccccc|}
\hline
\textbf{Class} &  0 - drawings & 1 - engraving & 2 - iconography & 3 - painting & 4 - sculpture\\
\hline
\textbf{Euclidean mAP} & 0.561 & 0.562 & 0.701 & 0.917 & 0.658 \\
\hline
\textbf{Cosine mAP} & 0.550 & 0.580 & 0.707 & 0.913 & 0.669 \\
\hline
\end{tabular}
\end{center}
\vspace{5mm}

Test introducing distractor
\begin{center}
\begin{tabular}{|l|cc|cc|c|}
\hline
k = 10 &  mAP euclidean & avg dist & mAP cosine & avg sim & $\#$Items\\
\hline
Resnet50 & 0.704 & 1.083 & 0.707 & 0.999 & 32791 \\
\hline
\end{tabular}
\end{center}

\begin{center}
\begin{tabular}{|l|ccccc|}
\hline
\textbf{Class} &  0 - drawings & 1 - engraving & 2 - iconography & 3 - painting & 4 - sculpture\\
\hline
\textbf{Euclidean mAP} & 0.549 & 0.554 & 0.690 & 0.914 & 0.639 \\
\hline
\textbf{Cosine mAP} & 0.536 & 0.572 & 0.697 & 0.909 & 0.648 \\
\hline
\end{tabular}
\end{center}
\vspace{2mm}
From the previous tables we can observe a very small decrease in the mAP values introducing the distractor both using the euclidean distance and the cosine similarity. \\

\subsubsection{LSH Index after Features Extraction, using Resnet50}
During this section we extracted the features from the dataset using the pre-trained network Resnet50 adopting a features extraction technique and we computed a kNN search setting w=4, obtaining the following results:
\begin{center}
\begin{tabular}{|l|cc|cc|c|}
\hline
k = 10 &  mAP euclidean & avg dist & mAP cosine & avg sim & $\#$Items\\
\hline
g=5 h=2 & 0,683 & 1.216 & 0,681 & 0.999  & 5701 \\
g=6 h=2 & 0,674 & 1.227 & 0,681 & 0.999  & 4836 \\
g=7 h=2 & 0,674 & 1.206 & 0,674 & 0.999  & 4872 \\
\textbf{g=8 h=2} & \textbf{0,681} & \textbf{1.241} & \textbf{0,687} & \textbf{0.998}  & \textbf{5304 }\\
g=7 h=6 & 0.583 & 1.518 & 0.589 & 0.998  & 1647 \\
g=4 h=5 & 0.556 & 1.491 & 0.630 & 0.998  & 922 \\
\hline
\end{tabular}
\end{center}
We tested the LSH index also with different values of \textit{w}, but we concluded that w=4 is the best values with Resnet50, because using lower values the buckets become sparse, while increasing \textit{w} the mAP remains constant.
\vspace{2mm}

\subsection{Resnet50v2}
\subsubsection{Sequential search after Features Extraction, using Resnet50v2}
During this section we extracted the features from the dataset using the pre-trained network Resnet50v2, adopting a features extraction technique and we computed a sequential scan \textit{both with and without the distractor}, in order to compare the performance when we introduced some images as noise.\\
We obtained the following results:\\

Test without distractor
\begin{center}
\begin{tabular}{|l|cc|cc|c|}
\hline
k = 10 &  mAP euclidean & avg dist & mAP cosine & avg sim & $\#$Items\\
\hline
Resnet50v2 & 0.893 & 17.986 & 0.902 & 0.747 & 7784 \\
\hline
\end{tabular}
\end{center}

\begin{center}
\begin{tabular}{|l|ccccc|}
\hline
\textbf{Class} &  0 - drawings & 1 - engraving & 2 - iconography & 3 - painting & 4 - sculpture\\
\hline
\textbf{Euclidean mAP} & 0.686 & 0.698 & 0.984 & 0.953 & 0.929 \\
\hline
\textbf{Cosine mAP} & 0.638 & 0.779 & 0.988 & 0.950 & 0.962 \\
\hline
\end{tabular}
\end{center}
\vspace{4mm}

Test introducing distractor
\begin{center}
\begin{tabular}{|l|cc|cc|c|}
\hline
k = 10 &  mAP euclidean & avg dist & mAP cosine & avg sim & $\#$Items\\
\hline
Resnet50v2 & 0.888 & 17.977 & 0.897 & 0.747 & 32791 \\
\hline
\end{tabular}
\end{center}

\begin{center}
\begin{tabular}{|l|ccccc|}
\hline
\textbf{Class} &  0 - drawings & 1 - engraving & 2 - iconography & 3 - painting & 4 - sculpture\\
\hline
\textbf{Euclidean mAP} & 0.682 & 0.696 & 0.984 & 0.947 & 0.918 \\
\hline
\textbf{Cosine mAP} & 0.636 & 0.772 & 0.988 & 0.946 & 0.953 \\
\hline
\end{tabular}
\end{center}
\vspace{3mm}

\noindent From the previous tables we can observe a very small decrease in the mAP values introducing the distractor both using the euclidean distance and the cosine similarity.\\
Moreover it is interesting to notice that using Resnet50v2 to extract the features from the images, the average distance computed using the euclidean distance between the objects is higher than the one computed using Resnet50 (about 17 times more), while the overall performance of the sequential scan search (measured by the mAP) is better, by about 18\%.\\

\subsubsection{LSH Index after Features Extraction, using Resnet50v2}
During this section we extracted the features from the dataset using the pre-trained network Resnet50v2, adopting a features extraction technique and we computed a kNN search, obtaining the following results:\\
\begin{center}
\begin{tabular}{|l|cc|cc|c|}
\hline
k = 10 &  mAP euclidean & avg dist & mAP cosine & avg sim & $\#$Items\\
\hline
\textbf{g=3 h=2} & \textbf{0.772} & \textbf{22.772} & \textbf{0.801} & \textbf{0.619}  & \textbf{477} \\
g=3 h=3 & 0.631 & 27.532 & 0.663 & 0.490   & 79 \\
g=3 h=4 & 0.357 & 29.709 & 0.404 & 0.409   & 25 \\ 
\hline
\textbf{g=4 h=2} & \textbf{0.793} & \textbf{22.285} & \textbf{0.814} & \textbf{0.634}  & \textbf{430} \\
g=4 h=3 & 0.608 & 27.865 & 0.651 & 0.474   & 71 \\
g=4 h=4 & 0.414 & 29.658 & 0.446 & 0.411   & 27 \\
\hline
\textbf{g=5 h=2} & \textbf{0.780} & \textbf{22.456} & \textbf{0.807} & \textbf{0.627}  & \textbf{464} \\
g=5 h=3 & 0.691 & 26.175 & 0.728 & 0.527   & 121 \\
g=5 h=4 & 0.486 & 28.903 & 0.509 & 0.432   & 23 \\
\hline
\textbf{g=6 h=2} & \textbf{0.794} & \textbf{21.995} & \textbf{0.822} & \textbf{0.639}  & \textbf{568} \\
g=6 h=3 & 0.677 & 26.748 & 0.705 & 0.513   & 88 \\
g=6 h=4 & 0.524 & 27.908 & 0.545 & 0.457   & 23 \\
\hline
\textbf{g=7 h=2} & \textbf{0.795} & \textbf{22.128} & \textbf{0.820} & \textbf{0.639}   & \textbf{685} \\
g=7 h=3 & 0.584 & 28.075 & 0.629 & 0.470   & 61 \\
g=7 h=4 & 0.304 & 29.794 & 0.323 & 0.407   & 22 \\
\hline
\textbf{g=8 h=2} & \textbf{0.789} & \textbf{22.137} & \textbf{0.818} & \textbf{0.635}   & \textbf{450} \\
g=8 h=3 & 0.566 & 27.591 & 0.615 & 0.483   & 74 \\
g=8 h=4 & 0.380 & 29.613 & 0.405 & 0.414   & 27 \\
\hline
\end{tabular}
\vspace{2mm}
\end{center}

\noindent We performed different tests varying the value of \textit{w}; in the previous table the values shown are calculated with \textbf{w=8}, indeed from other tests we noted a small decrease on the value of mAP, but not so relevant, so we reported only the best ones.\\

\subsection{Conclusion of the Features Extraction approach}
Observing the results obtained from the experiments based on features extraction approach, we can say that the distances between the features of the objects extracted with Resnet50v2 are greater than that obtained using Resnet50, as we expected. \\
Indeed passing an image as a query, the average euclidean distance between it and its k-nearest neighbor using Resnet50 is about $1.2$ while using Resnet50v2 is about $25$. Moreover the cosine similarity using Resnet50 is about $0.99$ while using Resnet50v2 is about $0.5$. \vspace{2mm}

\begin{figure}[h]
\centering
	\begin{minipage}[c]{.69\textwidth}
		\centering\setlength{\captionmargin}{0pt}%
		\includegraphics[width=1.\textwidth]{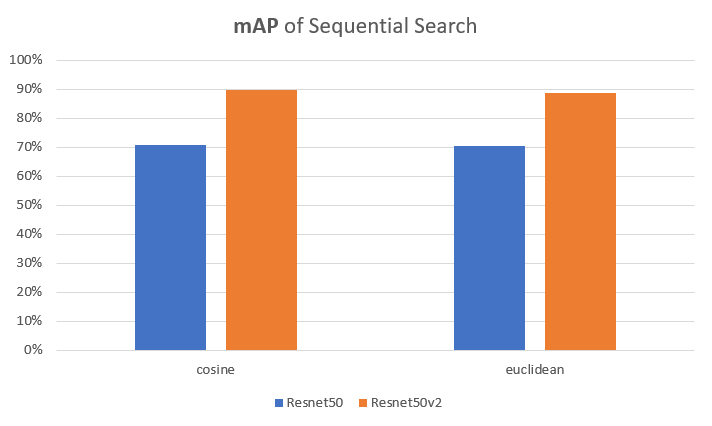}
		\label{fig:}
		\caption{Comparison of map value between Resnet50 and Resnet50v2}
	\end{minipage}
\end{figure}

\noindent We can see from the figure above that the difference between the Euclidean distance and the Cosine similarity in the search is minimal. In addition, while having features with a larger scale, the Resnet50v2 network has better performance than the Resnet50. \vspace{5mm} \\
In order to better compare the results obtained so far, we reported some other histograms.
The data included are both from the sequential search and the LSH index search. For the latter, the hyper-parameters taken for the comparison are:
\begin{itemize}
\item Resnet50: $g=8, h=2, w=4$
\item Resnet50v2: $g=7, h=2, w=8$
\end{itemize}

If we look at the sequential search performance, Resnet50v2 gets worse by 8\% while Resnet50 only worsens by 2\%. However, we can see that Resnet50v2 has a better mAP even with the index, with 82\% compared to 69\% of Resnet50. We also have to point out that the average number of objects visited per query by Resnet50v2 is 8 times lower than Resnet50, and this might be the reason why there is so much difference in the mAP from sequential to LSH of the first mentioned network. 

\begin{figure}[H]
\centering
	\begin{minipage}[c]{.45\textwidth}
		\centering\setlength{\captionmargin}{0pt}%
		\includegraphics[width=1.\textwidth]{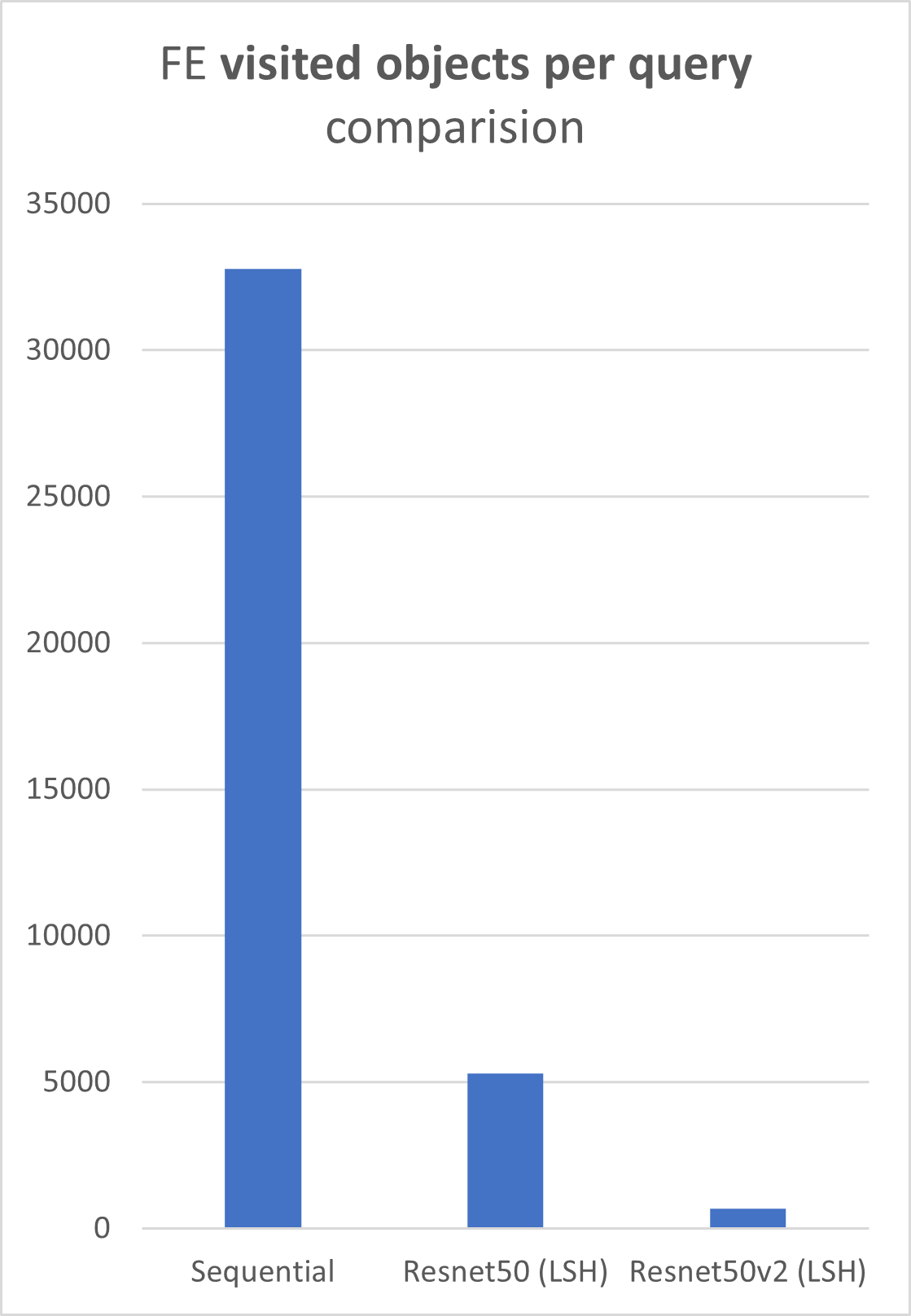}
		\label{fig:}
	\end{minipage}
	\begin{minipage}[c]{.54\textwidth}
		\centering\setlength{\captionmargin}{0pt}%
		\includegraphics[width=1.\textwidth]{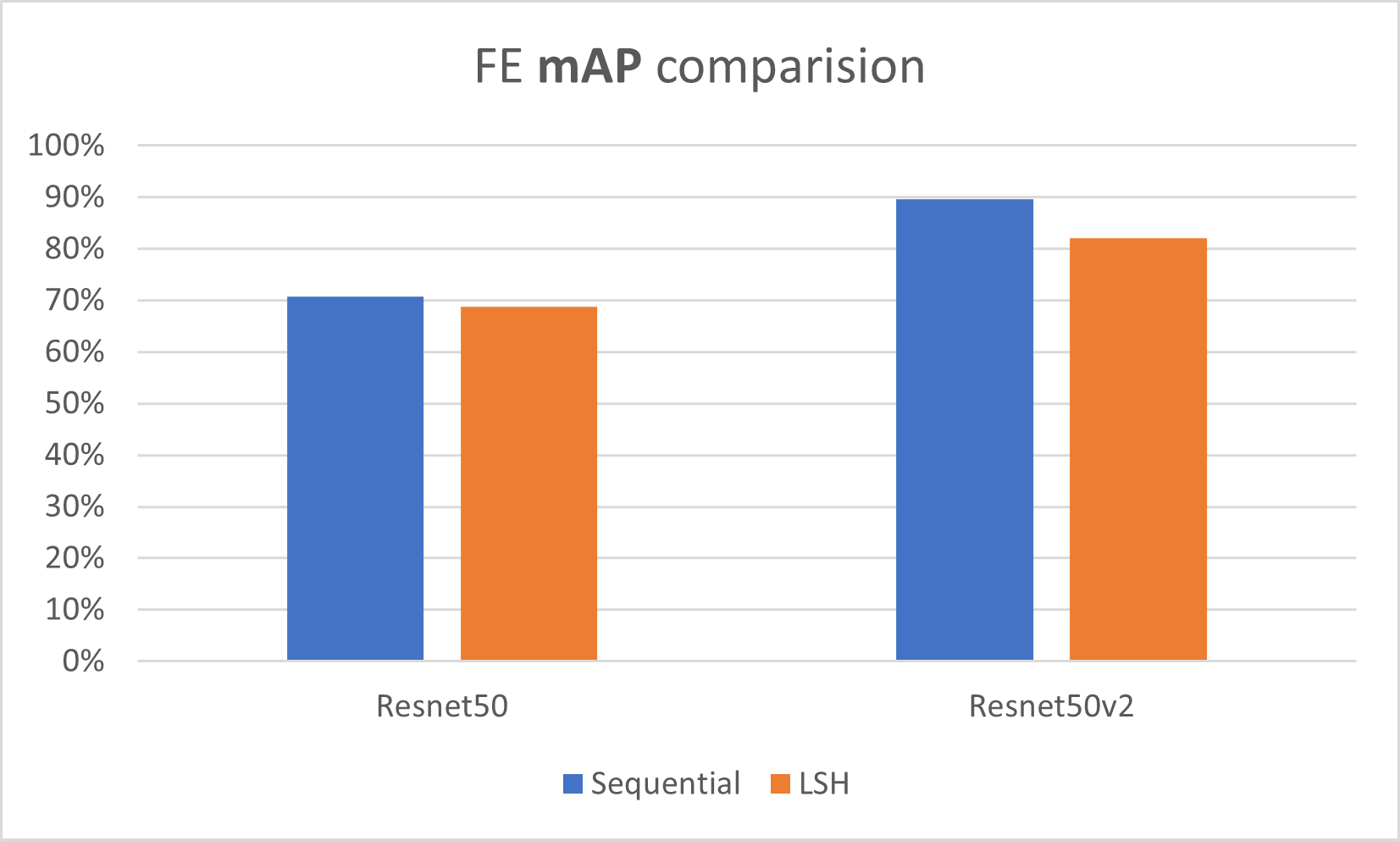}
		\label{fig:}
		\centering\setlength{\captionmargin}{0pt}%
		\includegraphics[width=1.\textwidth]{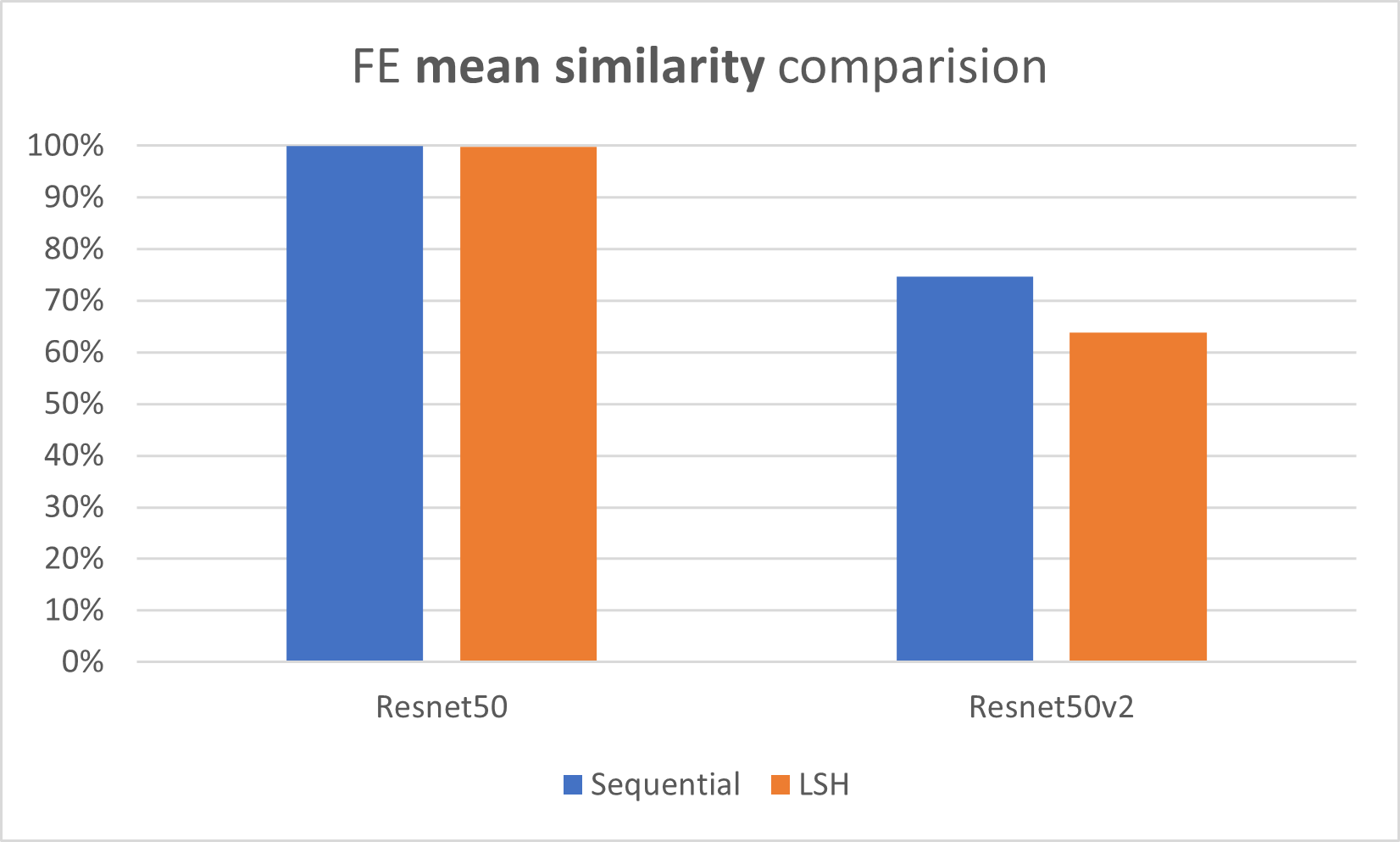}
		\label{fig:}
	\end{minipage}%
	\caption{Data comparision for the Feature Extraction approach}
\end{figure}

\section{Fine Tuning approach}\label{sec:5}
For the Fine-Tuning phase, we carried out several tests using different hyper-parameters for the fully connected part added at the end of both the networks Resnet50 and Resnet50v2.
We stopped the training exploiting an \textit{early stopping condition}, monitoring the loss function and setting the patience value equal to 10 and restoring the weights to 10 epochs before. The following paragraphs show only the most relevant experiments.\\
\subsection{Resnet50}

\subsubsection{Model 1}
The first model has been obtained using ResNet50 as base network, unfreezing from layer \textit{conv5\_block1\_1\_conv}; the fully connected network is made up of three pairs of levels, each of which is composed of a dropout layer (with drop rate of 0.5) and a fully connected layer (with 256, 128 and 64 neurons). The activation function chosen is "softmax"; the learning rate has been set to 1e-6. Below are the evaluation metrics obtained from the network training.

\begin{verbatim}

        class    precision  recall   f1-score   support

         0.0       0.71      0.34      0.46       122
         1.0       0.64      0.68      0.66        84
         2.0       0.80      0.95      0.87       231
         3.0       0.96      0.87      0.91       228
         4.0       0.76      0.92      0.83       191

    accuracy                           0.81       856
   macro avg       0.78      0.75      0.75       856
weighted avg       0.81      0.81      0.79       856

loss_test : 0.6162440776824951     
acc_test : 0.8072429895401001

\end{verbatim}
\begin{figure}[H]
    \centering
    \begin{minipage}{0.5\textwidth}
        \centering
        \includegraphics[scale=0.46]{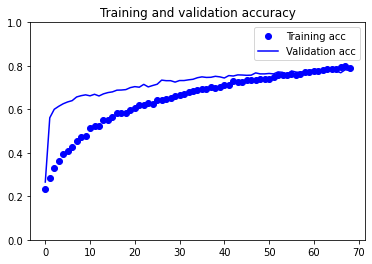}
    \end{minipage}\hfill
    \begin{minipage}{0.5\textwidth}
        \centering
        \includegraphics[scale=0.46]{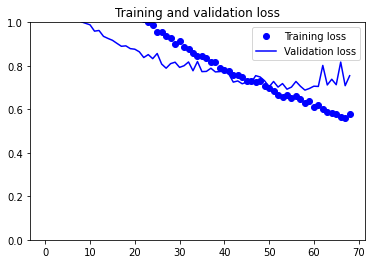}
    \end{minipage}
    \caption{Accuracy and loss of Model 1 - Resnet50}
\end{figure}

\noindent The model above obtained the best results in terms of accuracy and loss on the validation set ($81\%$ and $61\%$, respectively) with respect to all hyper-parameter configurations tested.

\subsubsection{Model 2}
For the second model, the same fully connected network of the previous experiment has been appended to the regular ResNet50 network; a data augmentation technique has been applied in order to better generalize the dataset during the training.

\begin{verbatim}

        class    precision  recall   f1-score   support

         0.0       0.62      0.28      0.38       122
         1.0       0.53      0.74      0.62        84
         2.0       0.85      0.87      0.86       231
         3.0       0.91      0.85      0.88       228
         4.0       0.71      0.87      0.79       191

    accuracy                           0.77       856
   macro avg       0.72      0.72      0.70       856
weighted avg       0.77      0.77      0.76       856

loss_test : 0.6571111083030701
acc_test : 0.7675233483314514

\end{verbatim}

\begin{figure}[H]
    \centering
    \begin{minipage}{0.5\textwidth}
        \centering
        \includegraphics[scale=0.46]{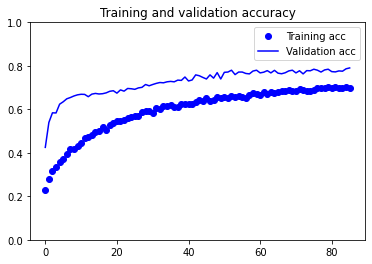}
    \end{minipage}\hfill
    \begin{minipage}{0.5\textwidth}
        \centering
        \includegraphics[scale=0.46]{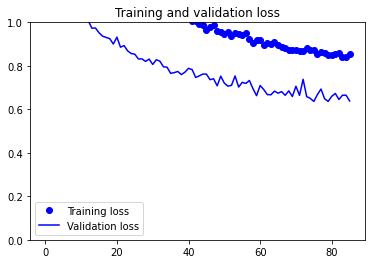}
    \end{minipage}
    \caption{Accuracy and loss of Model 2 - Resnet50}
\end{figure}
\vspace{3mm}
\noindent The performance did not improve, in fact a slight underfitting appeared.

\subsubsection{Sequential search after Fine Tuning, using Resnet50}
During this section we extracted the features from the dataset using the pre-trained network Resnet50, adopting a fine tuning technique and we computed a sequential scan, obtaining the following results:
\begin{center}
\begin{tabular}{|l|cc|cc|cc|c|}
\hline
 & \textbf{Accuracy} & \textbf{Loss} & \textbf{mAP} euclidean & avg dist & \textbf{mAP} cosine & avg sim & Items\\
\hline
\textbf{Model 1} & \textbf{0.8072} & \textbf{0.6162} &  \textbf{0.767} & \textbf{4.413} & \textbf{0.780} & \textbf{0.943} & \textbf{32791}\\
Model 2 & 0.7675 & 0.6571 &  0.756 & 3.679 & 0.768 & 0.953 &32791\\
\hline
\end{tabular}
\end{center}

\subsubsection{LSH Index after Fine Tuning, using Resnet50}
During this section we extracted the features from the dataset using the pre-trained network Resnet50, adopting a fine tuning technique and we computed an indexed kNN search, obtaining the following results (w=4):
\begin{center}
\begin{tabular}{|l|cc|cc|c|}
\hline
  & \textbf{mAP} euclidean & avg dist & \textbf{mAP} cosine & avg sim & Items\\
\hline
\textbf{Model 1, g=5 h=2} &  \textbf{0.711} & \textbf{4.908} & \textbf{0.733} & \textbf{0.931} & \textbf{1452} \\
Model 2, g=5 h=2 &  0.707 & 4.011 & 0.728 & 0.944 & 2397\\
\hline
\end{tabular}
\vspace{1mm}
\end{center}

\noindent Comparing the performance obtained using different models, we can conclude that computing a sequential scan or using the index, the performance change basically only in terms of items retrieved. Moreover we reported only the best results of the indexes: selected a model and varying the hyper-parameters \textit{g} and \textit{h} the indexes that retrieve less objects. Finally we chose the Model 1 with the following hyper-parameters g=5, h=2, w=4 because it visits less objects per query with respect to the others, while maintaining a good value of mAP. \vspace{5mm} \\
Below there are some histograms summarizing the results obtained with the Resnet50 network. The index data are associated to the hyper-parameters  that gave the best performance in both the Feature Extraction ($g=8, h=2, w=4$) and Fine Tuning ($g=5, h=2, w=4$) approaches. Moreover, for the latter approach we refer to the Model 1 results. \\
After the re-training of the last blocks of the network on our training set, we can see that the mAP has increased to 78\% in the sequential search, and to 73\% in the indexed search. Not only we have an increase in the mAP, but there is also a large decrease in the average number of objects visited per query, from 5304 to 1452. \\
Both of these considerations lead us to conclude that this approach achieved the best index search with the Resnet50 network.

\begin{figure}[H]
\centering
	\begin{minipage}[c]{.4\textwidth}
		\centering\setlength{\captionmargin}{0pt}%
		\includegraphics[width=1.\textwidth]{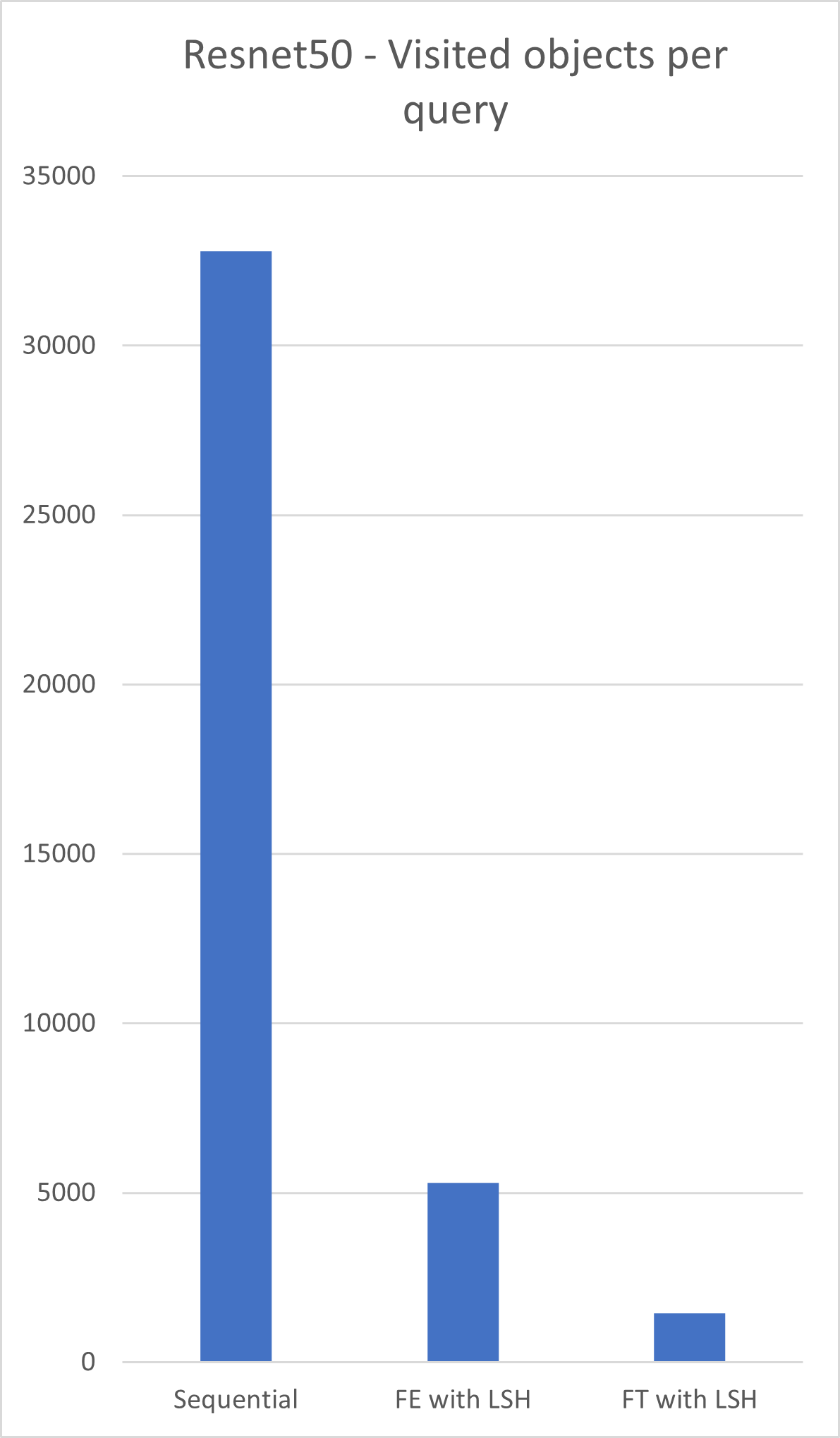}
		\label{fig:}
	\end{minipage}
	\begin{minipage}[c]{.57\textwidth}
		\centering\setlength{\captionmargin}{0pt}%
		\includegraphics[width=1.\textwidth]{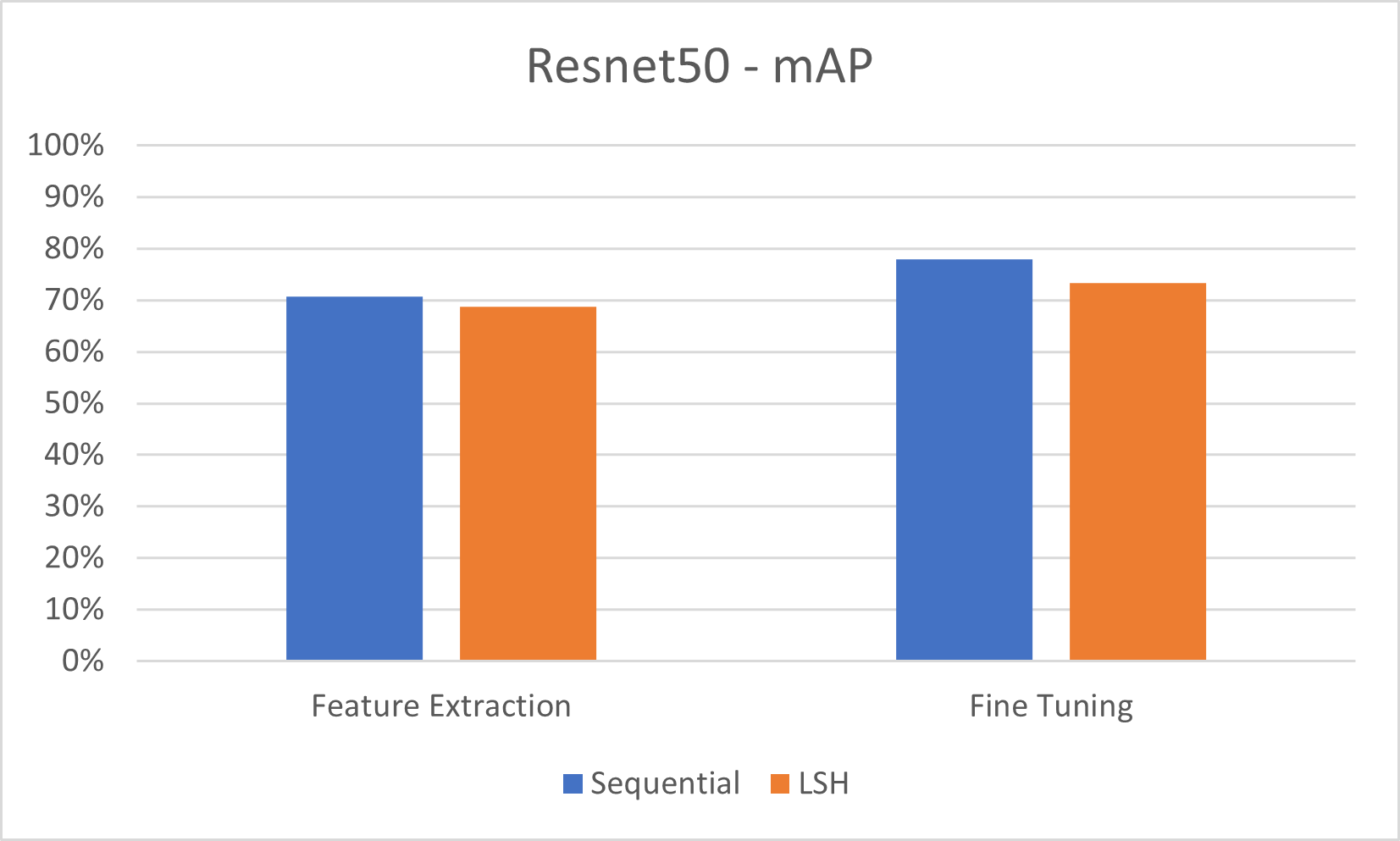}
		\label{fig:}
		\centering\setlength{\captionmargin}{0pt}%
		\includegraphics[width=1.\textwidth]{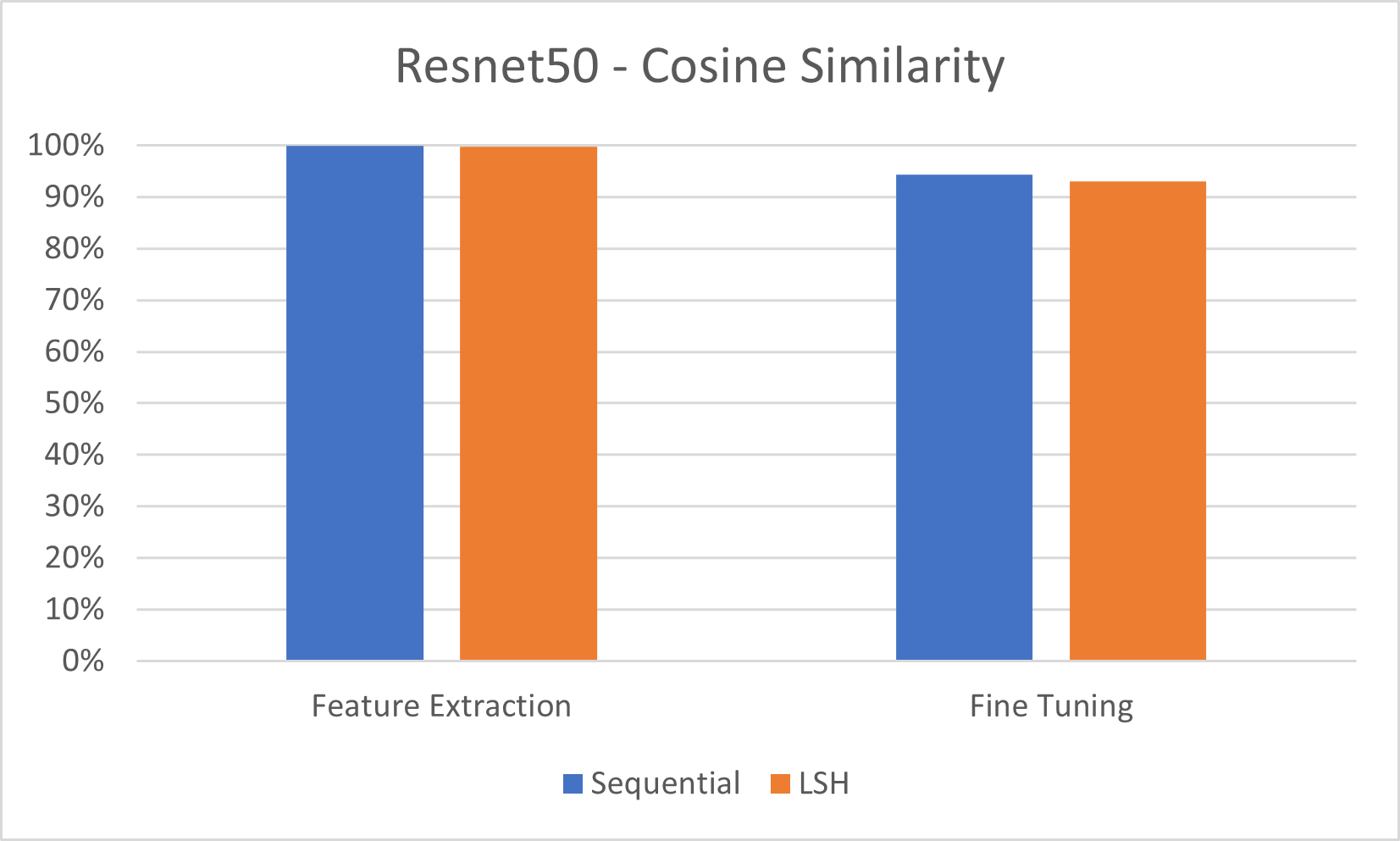}
		\label{fig:}
	\end{minipage}%
	\caption{Final data comparision for the Resnet50 network}
\end{figure}
\vspace{3mm}

\subsection{Resnet50v2}
For this convolutional base, we performed the tests by resuming the fully connected blocks used in the previous step. We just report the obtained results, since all the hyper-parameters can be consulted in the previous section.

\subsubsection{Model 1}
The first model has been obtained using ResNet50V2 as base network, unfreezing from layer \textit{conv5\_block1\_1\_conv}; the fully connected network is made up of three pairs of levels, each of which is composed of a dropout layer (with drop rate of 0.5) and a fully connected layer (with 256, 128 and 64 neurons). The activation function chosen is "softmax"; the learning rate has been set to 1e-6. Below are the evaluation metrics obtained from the network training.

\begin{verbatim}
        class    precision  recall   f1-score   support

         0.0       0.90      0.21      0.34       122
         1.0       0.61      0.83      0.70        84
         2.0       0.97      0.98      0.98       231
         3.0       0.83      0.96      0.89       228
         4.0       0.87      0.97      0.92       191

    accuracy                           0.85       856
   macro avg       0.83      0.79      0.77       856
weighted avg       0.86      0.85      0.82       856

loss_test : 0.48188671469688416  
acc_test : 0.8492990732192993
\end{verbatim}

\begin{figure}[H]
    \centering
    \begin{minipage}{0.5\textwidth}
        \centering
        \includegraphics[scale=0.46]{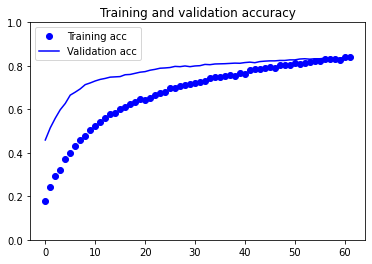}
    \end{minipage}\hfill
    \begin{minipage}{0.5\textwidth}
        \centering
        \includegraphics[scale=0.46]{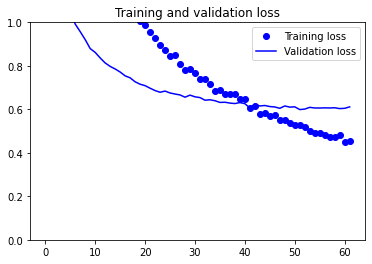}
    \end{minipage}
    \caption{Accuracy and loss of Model 1 - Resnet50v2}
\end{figure}
\vspace{3mm}

\subsubsection{Model 2}
For the second model, the same fully connected network of the previous experiment has been appended to the regular ResNet50V2 network; a data augmentation technique has been applied in order to better generalize the dataset during the training.

\begin{verbatim}
        class    precision  recall   f1-score   support

         0.0       0.78      0.48      0.60       122
         1.0       0.70      0.73      0.71        84
         2.0       0.96      0.99      0.98       231
         3.0       0.90      0.95      0.93       228
         4.0       0.88      0.99      0.93       191

    accuracy                           0.88       856
   macro avg       0.84      0.83      0.83       856
weighted avg       0.88      0.88      0.87       856

loss_test : 0.37721073627471924  
acc_test : 0.8808411359786987

\end{verbatim}

\begin{figure}[H]
    \centering
    \begin{minipage}{0.5\textwidth}
        \centering
        \includegraphics[scale=0.46]{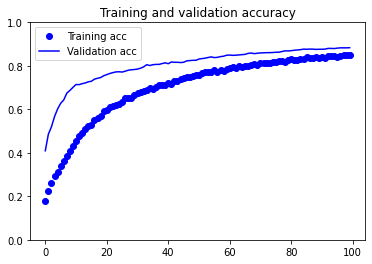}
    \end{minipage}\hfill
    \begin{minipage}{0.5\textwidth}
        \centering
        \includegraphics[scale=0.46]{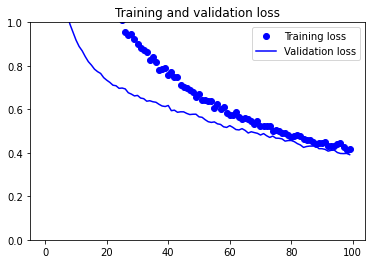}
    \end{minipage}
    \caption{Accuracy and loss of Model 2 - Resnet50v2}
\end{figure}
\vspace{3mm}
\noindent In this case the performance has improved with respect to the first model, accuracy increased from $85\%$ to $88\%$ and loss decreased passing from $48\%$ to $38\%$.

\subsubsection{Sequential search after Fine Tuning Resnet50v2}
During this section we extracted the features from the dataset using the pre-trained network Resnet50v2 after having fine tuned it and we computed a sequential scan, obtaining the following results:
\begin{center}
\begin{tabular}{|l|cc|cc|cc|c|}
\hline
 & \textbf{Accuracy} & \textbf{Loss} & \textbf{mAP} euclidean & avg dist & \textbf{mAP} cosine & avg sim & Items\\
\hline
Model 1 & 0.8493 & 0.4819 &  0.875 & 16.684 & 0.886 & 0.661 & 32791\\
\textbf{Model 2} & \textbf{0.8808} & \textbf{0.3772} &  \textbf{0.893} &  \textbf{8.997} & \textbf{0.908} & \textbf{0.729} & \textbf{32791}\\
\hline
\end{tabular}
\end{center}

\subsubsection{LSH Index after Fine Tuning Resnet50v2}
During this section we extracted the features from the dataset using the pre-trained network Resnet50v2, adopting a fine tuning technique and we computed an indexed kNN search, obtaining the results in the table below. \vspace{3mm} \\
We finally chose the Model 2 with the hyper-parameters $g=7, h=2, w=6$ because it has the higher value of mAP and it still visits a low number of objects.

\begin{center}
\begin{tabular}{|l|l|cc|cc|c|}
\hline
   & Hyperparameters & \textbf{mAP} euclidean & avg dist & \textbf{mAP} cosine & avg sim & Items\\
\hline
Model 1 & g=4 h=2 w=5 &  0.725 & 22.231 & 0.793 & 0.478 & 266\\
Model 1 & g=4 h=2 w=6 &  0.764 & 22.038 & 0.824 & 0.516 & 407\\
\hline
Model 1 & g=5 h=2 w=5 &  0.738 & 21.862 & 0.800 & 0.489 & 284\\
Model 1 & g=5 h=2 w=6 &  0.760 & 21.431 & 0.822 & 0.504 & 423\\
\hline
Model 1 & g=6 h=2 w=5 &  0.710 & 21.759 & 0.801 & 0.486 & 254\\
Model 1 & g=6 h=2 w=6 &  0.752 & 20.978 & 0.819 & 0.511 & 369\\
\hline
Model 1 & g=7 h=2 w=5 &  0.721 & 21.851 & 0.799 & 0.488 & 345\\
\textbf{Model 1} & \textbf{g=7 h=2 w=6} &  \textbf{0.761} & \textbf{20.861} & \textbf{0.830} & \textbf{0.519} & \textbf{532} \\
\hline
\hline
Model 2 & g=4 h=2 w=5 &  0.790 & 11.181 & 0.840 & 0.611 & 724\\
Model 2 & g=4 h=2 w=6 &  0.793 & 11.661 & 0.840 & 0.596 & 727\\
\hline
Model 2 & g=5 h=2 w=5 &  0.808 & 11.147 & 0.848 & 0.607 & 1061\\
Model 2 & g=5 h=2 w=6 &  0.823 & 10.958 & 0.865 & 0.625 & 1534\\
\hline
Model 2 & g=6 h=2 w=5 &  0.797 & 10.887 & 0.847 & 0.616 & 704\\
Model 2 & g=6 h=2 w=6 &  0.777 & 10.958 & 0.838 & 0.616 & 1012\\
\hline
Model 2 & g=7 h=2 w=5 &  0.814 & 11.293 & 0.863 & 0.606 & 1007\\
\textbf{Model 2} & \textbf{g=7 h=2 w=6} &  \textbf{0.829} & \textbf{11.002} & \textbf{0.861} & \textbf{0.623} & \textbf{1440} \\
\hline
\end{tabular}
\end{center}

\vspace{3mm} 
\noindent Below there are some histograms summarizing the results obtained with the Resnet50v2 network. The index data are associated to the hyper-parameters  that gave the best performance in both the Feature Extraction ($g=7, h=2, w=8$) and Fine Tuning ($g=7, h=2, w=6$) approaches. Moreover, for the latter approach we refer to the Model 2 results. \\
\begin{figure}[H]
\centering
	\begin{minipage}[c]{.41\textwidth}
		\centering\setlength{\captionmargin}{0pt}%
		\includegraphics[width=1.\textwidth]{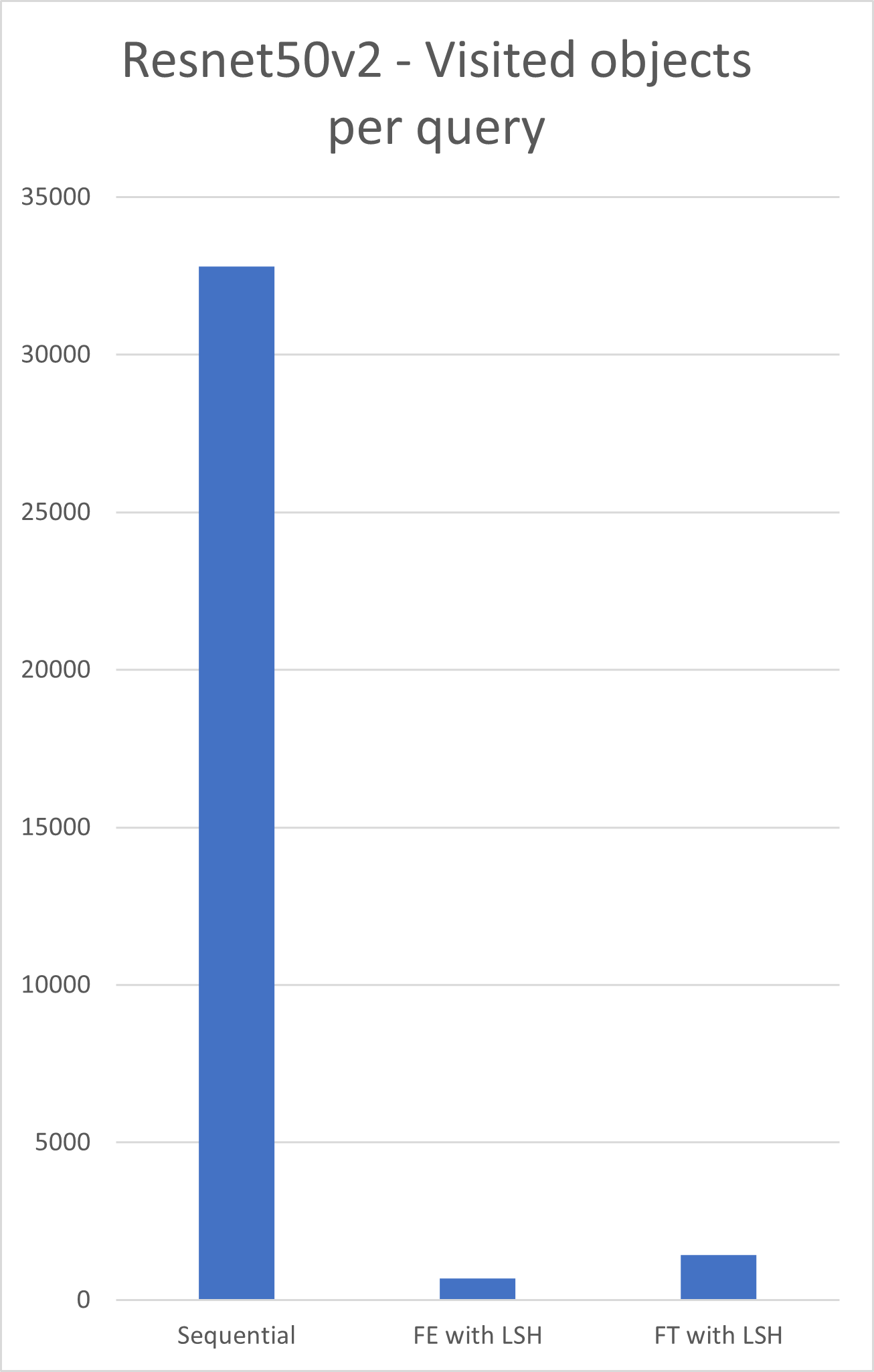}
		\label{fig:}
	\end{minipage}
	\begin{minipage}[c]{.54\textwidth}
		\centering\setlength{\captionmargin}{0pt}%
		\includegraphics[width=1.\textwidth]{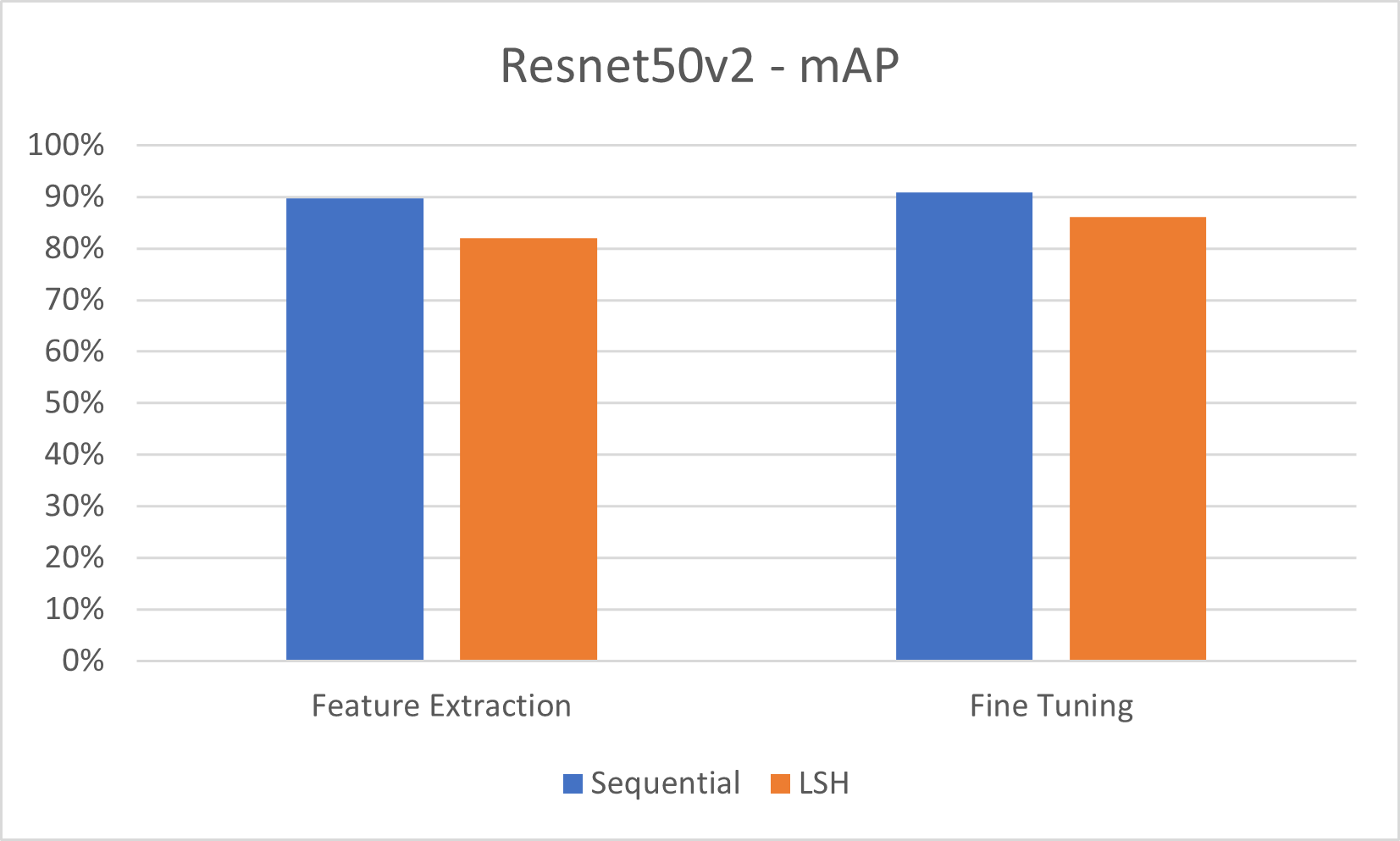}
		\label{fig:}
		\centering\setlength{\captionmargin}{0pt}%
		\includegraphics[width=1.\textwidth]{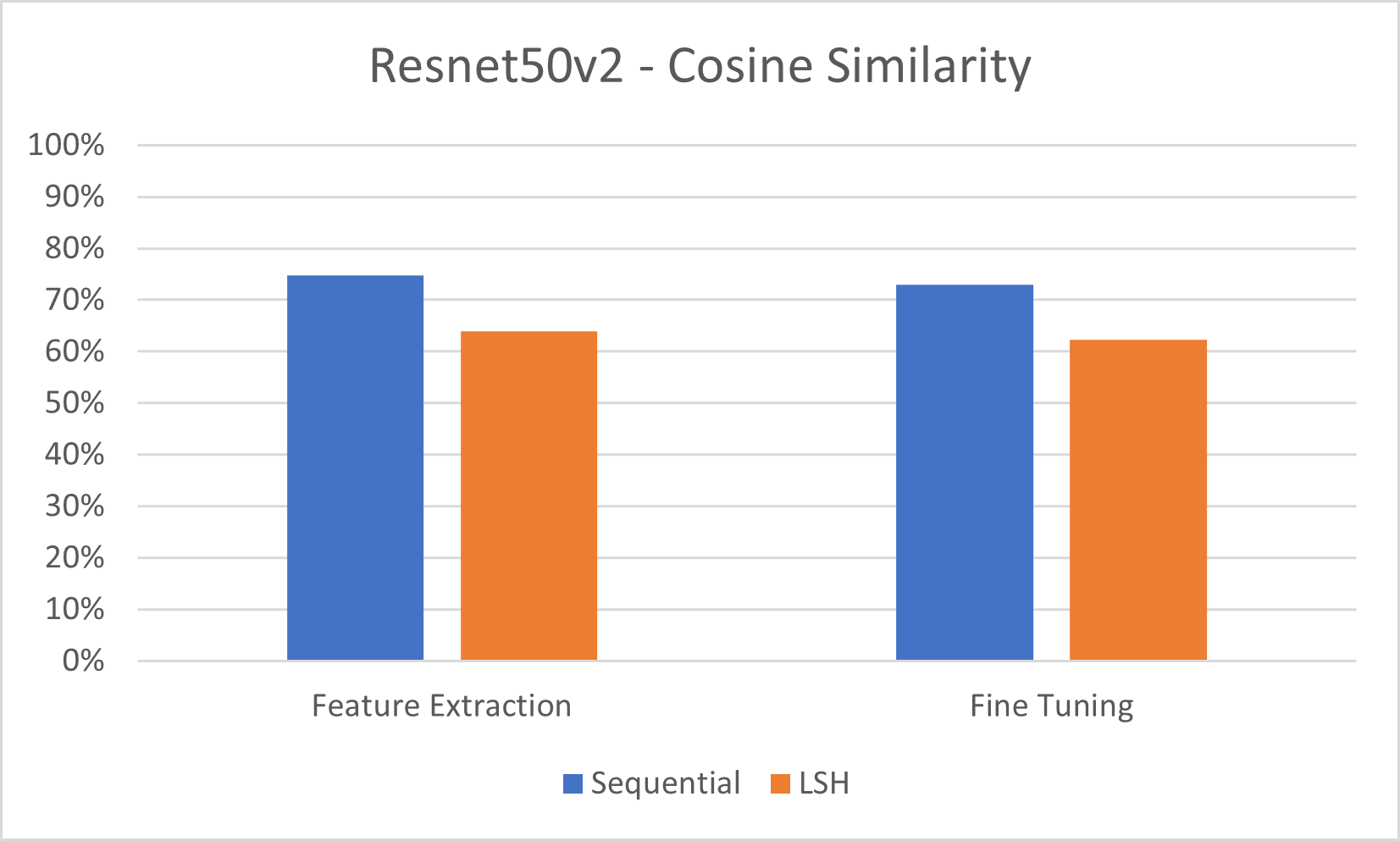}
		\label{fig:}
	\end{minipage}%
	\caption{Final data comparision for the Resnet50v2 network}
\end{figure}

\noindent In this case, the partial re-training of the network brought the mAP to 86\% using the index, whereas before it was 82\%. This value of mAP is the highest obtained so far with the LSH index search. The number of visited objects is raised compared to the feature extraction data, from 685 to 1440, still remaining lower than the Resnet50 data.\\ 
This is given by the fact that the fine tuning technique has made the features closer than features extraction: indeed in the first approach we selected $w=8$ in the index creation, while in the second approach $w=6$ was enough to obtain a high average similarity and a high mAP.

\vspace{3mm}

\subsection{Conclusion of Fine Tuning approach}
In order to better compare the obtained results, we reported some other histograms. \\
The data included are both from the sequential search and the LSH index search. For the latter, the hyper-parameters taken for the comparison are:
\begin{itemize}
\item Resnet50 (model 2): $g=5, h=2, w=4$
\item Resnet50v2 (model 2): $g=7, h=2, w=6$
\end{itemize}
\vspace{3mm}
We see that, for the same average objects visited per query, the model built on Resnet50v2 offers much higher performance, with a mAP of 86\%, versus Resnet50's 73\%. 
\begin{figure}[H]
\centering
	\begin{minipage}[c]{.45\textwidth}
		\centering\setlength{\captionmargin}{0pt}%
		\includegraphics[width=1.\textwidth]{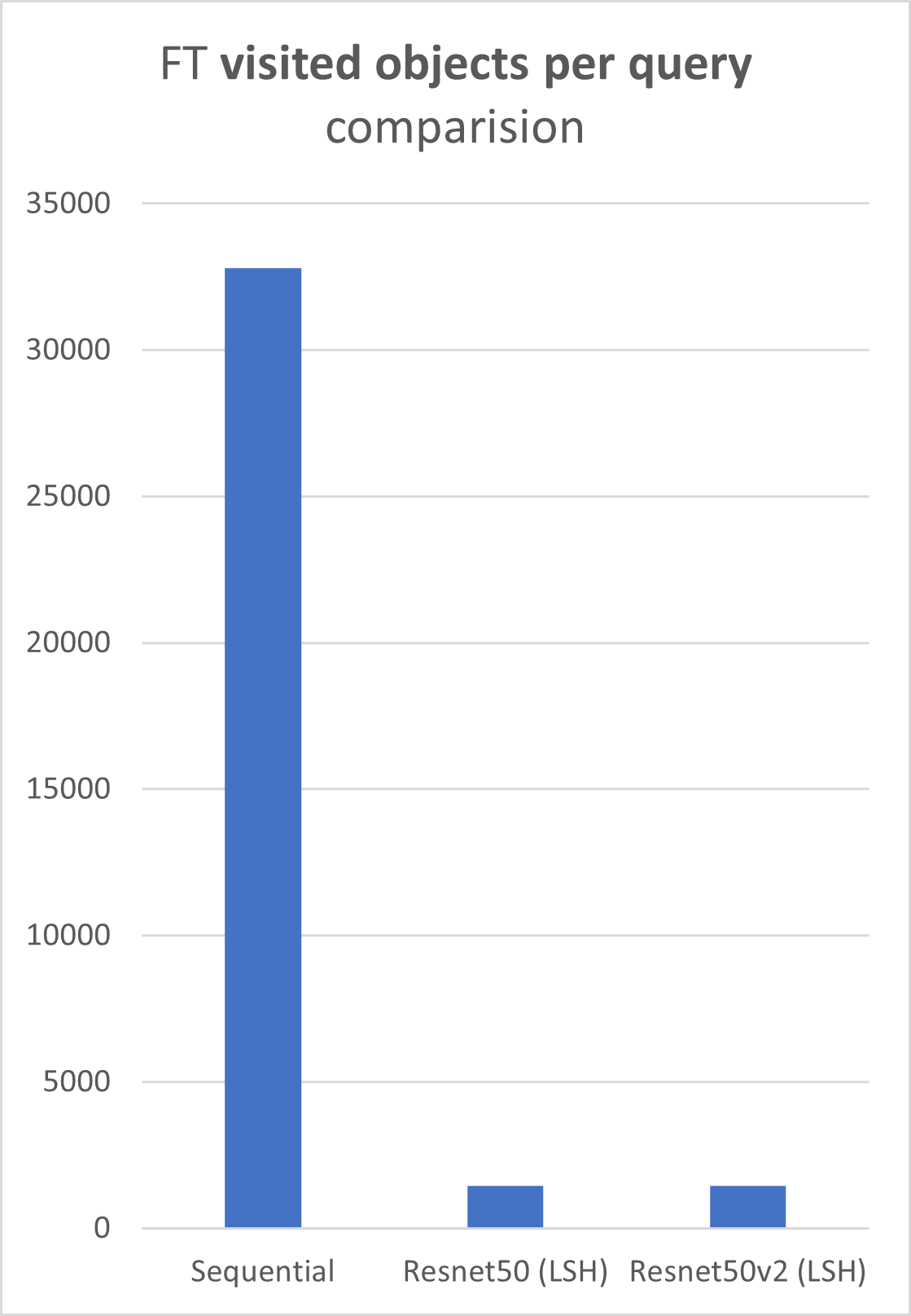}
		\label{fig:}
	\end{minipage}
	\begin{minipage}[c]{.54\textwidth}
		\centering\setlength{\captionmargin}{0pt}%
		\includegraphics[width=1.\textwidth]{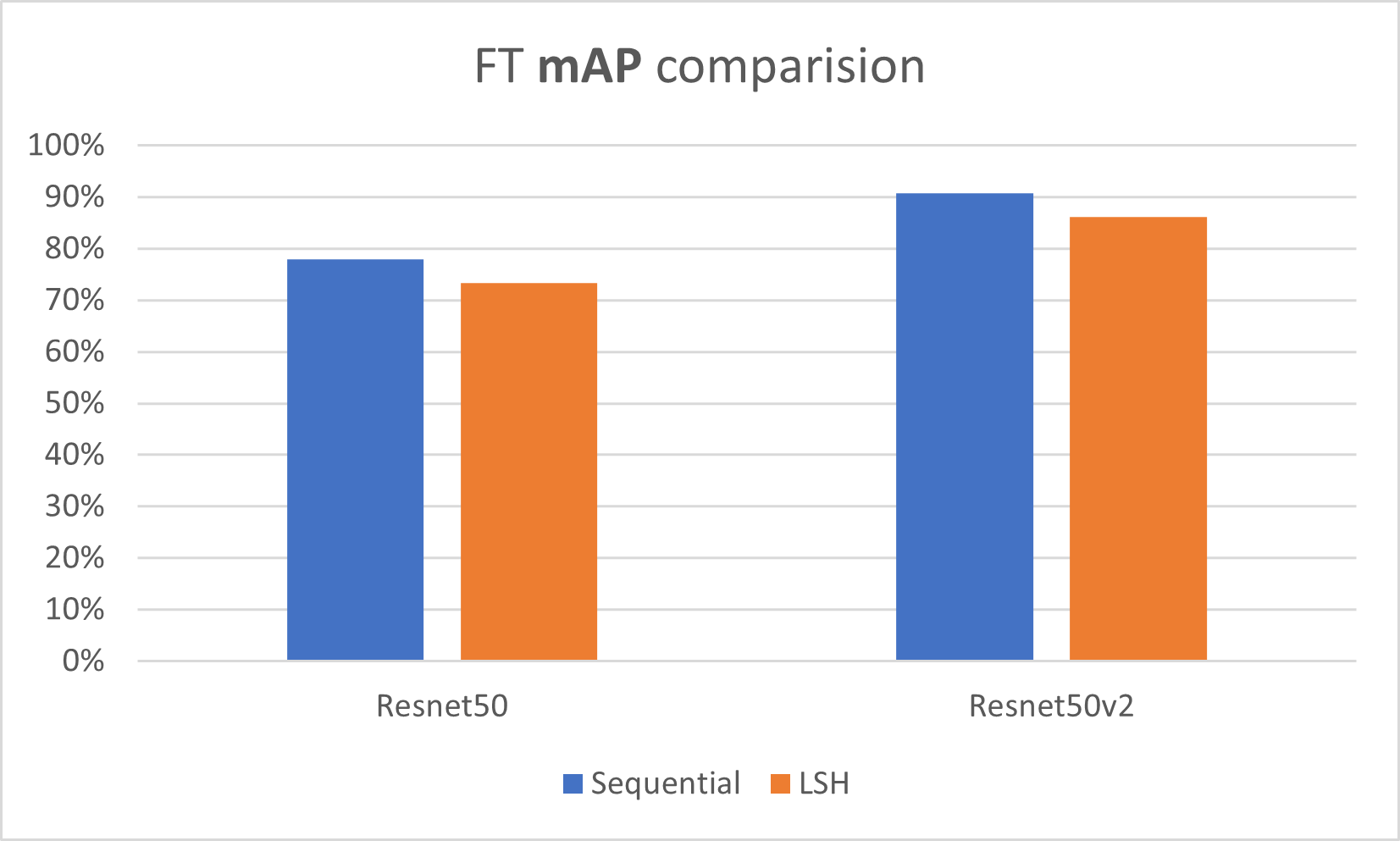}
		\label{fig:}
		\centering\setlength{\captionmargin}{0pt}%
		\includegraphics[width=1.\textwidth]{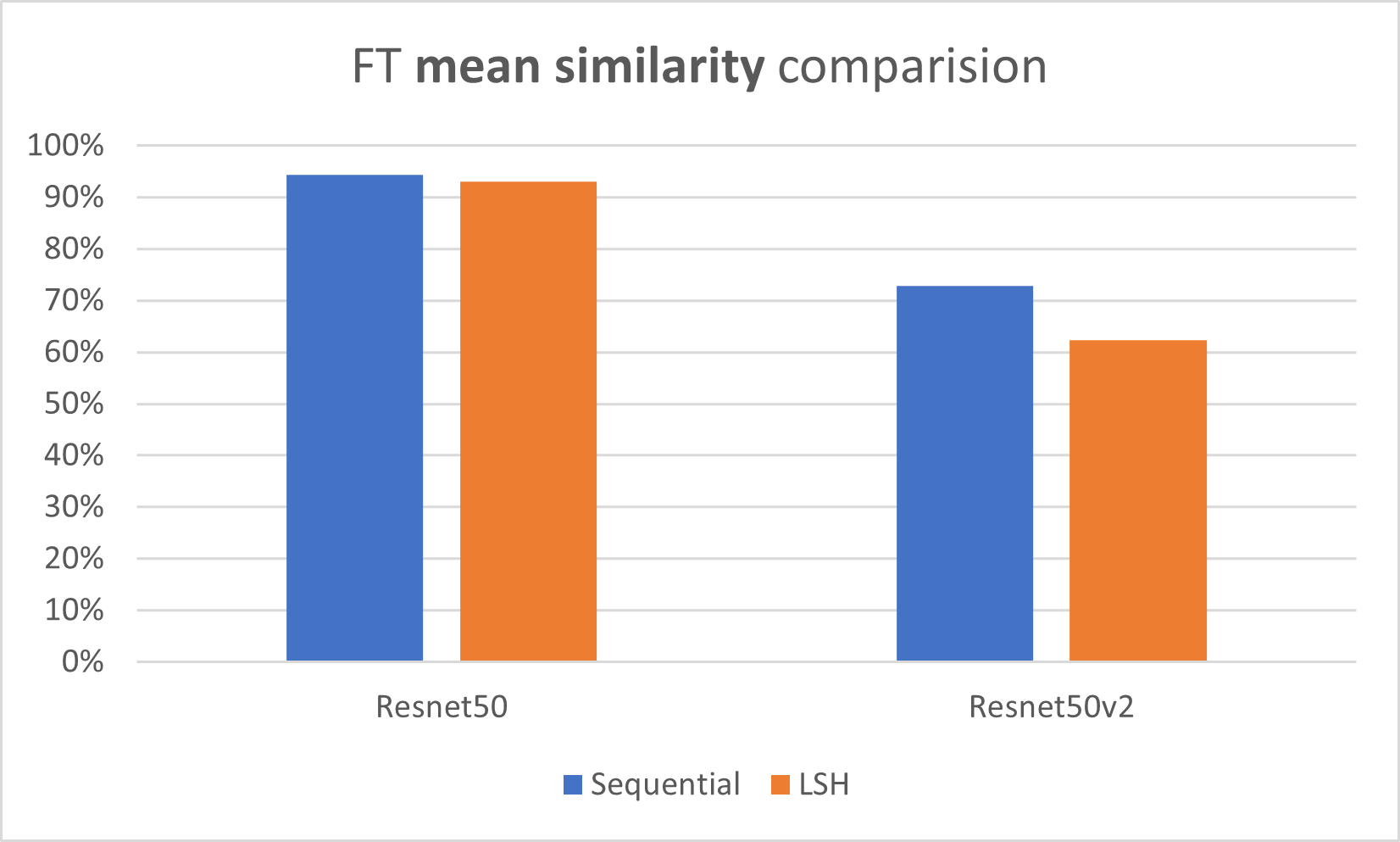}
		\label{fig:}
	\end{minipage}%
	\caption{Data comparision for the Fine Tuning approach}
\end{figure}
\vspace{2mm}

\section{LSH index considerations}\label{sec:6}
During this part of the project we analyzed the space of the hyper-parameters values after running different tests in the previous sections. All the following figures show plots where the data are computed using the cosine similarity, but similar results are obtained using the euclidean distance.\\

\noindent The first aspect that we noted is that we obtained the highest values of the mAP setting a low value of the number of hyper-planes \textit{h} (h=2) fixed the length of the segment \textit{w} and the number of g-functions \textit{g}: indeed increasing the value of \textit{h} we can observe lower curves. \\
Moreover we can observe that increasing the value of \textit{w}, the value of mAP also increases.
\begin{figure}[H]
    \centering
    \begin{minipage}{0.5\textwidth}
        \includegraphics[width=1.\textwidth]{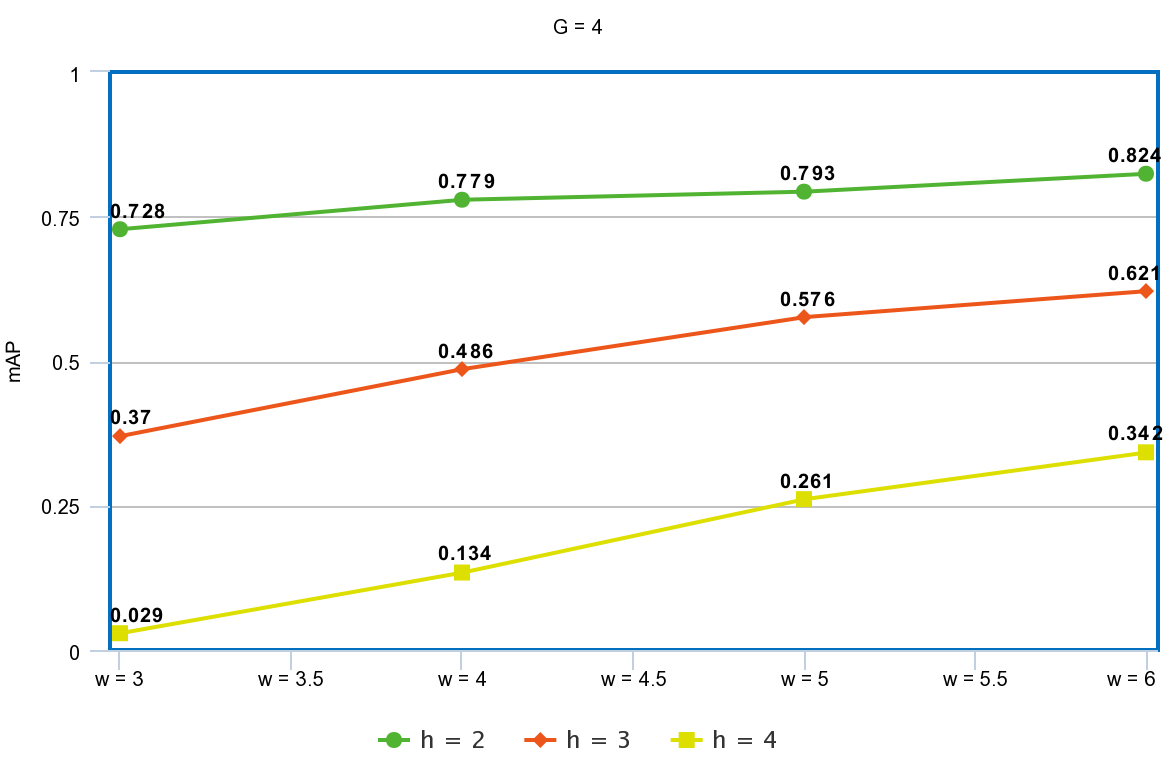}
    \end{minipage}
    \caption{Resnet50v2 fine tuned model 2 - \textbf{mAP} trend with fixed $g=4$}
\end{figure}

\noindent The following two figures show the variation of the mAP value and the number of visited objects when we increase the number of g-functions.
\begin{figure}[H]
    \centering
    \begin{minipage}{0.49\textwidth}
        \centering
        \includegraphics[width=1.\textwidth]{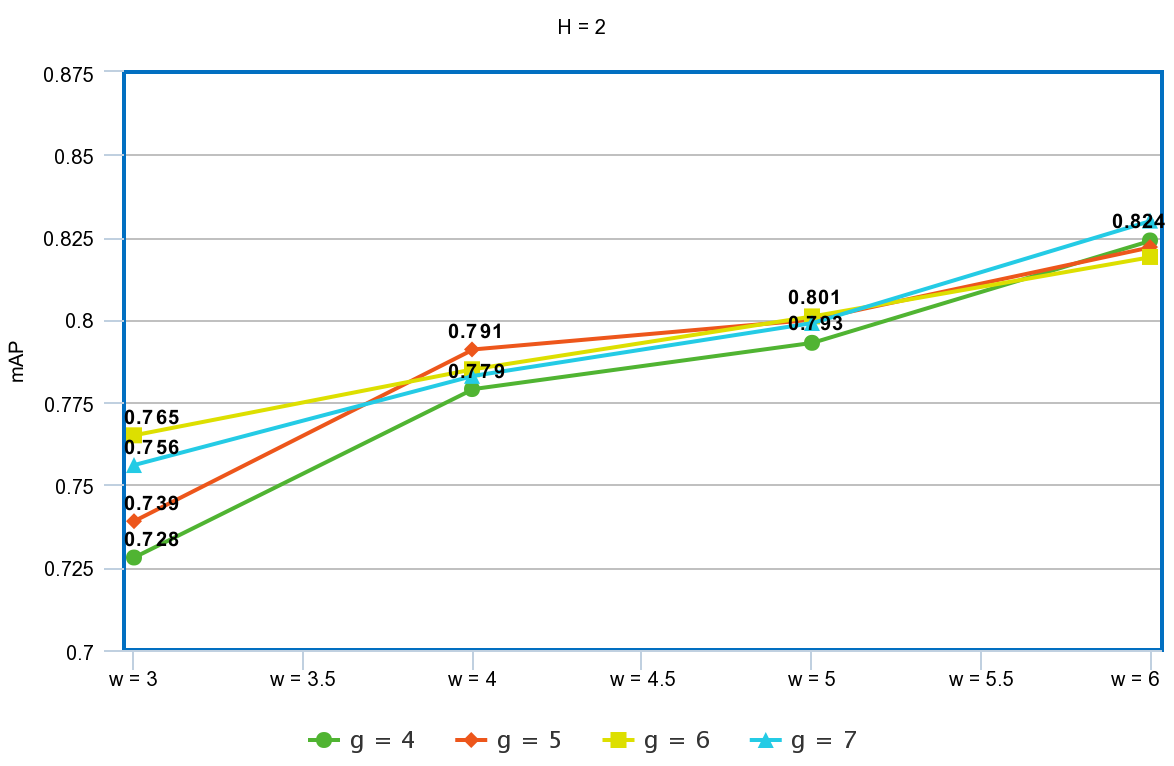}
        \caption{Resnet50v2 fine tuned model 2 - \textbf{mAP} trend with fixed $h=2$}
    \end{minipage}\hfill
    \hspace{1mm}
    \begin{minipage}{0.49\textwidth}
        \centering
        \includegraphics[width=1.\textwidth]{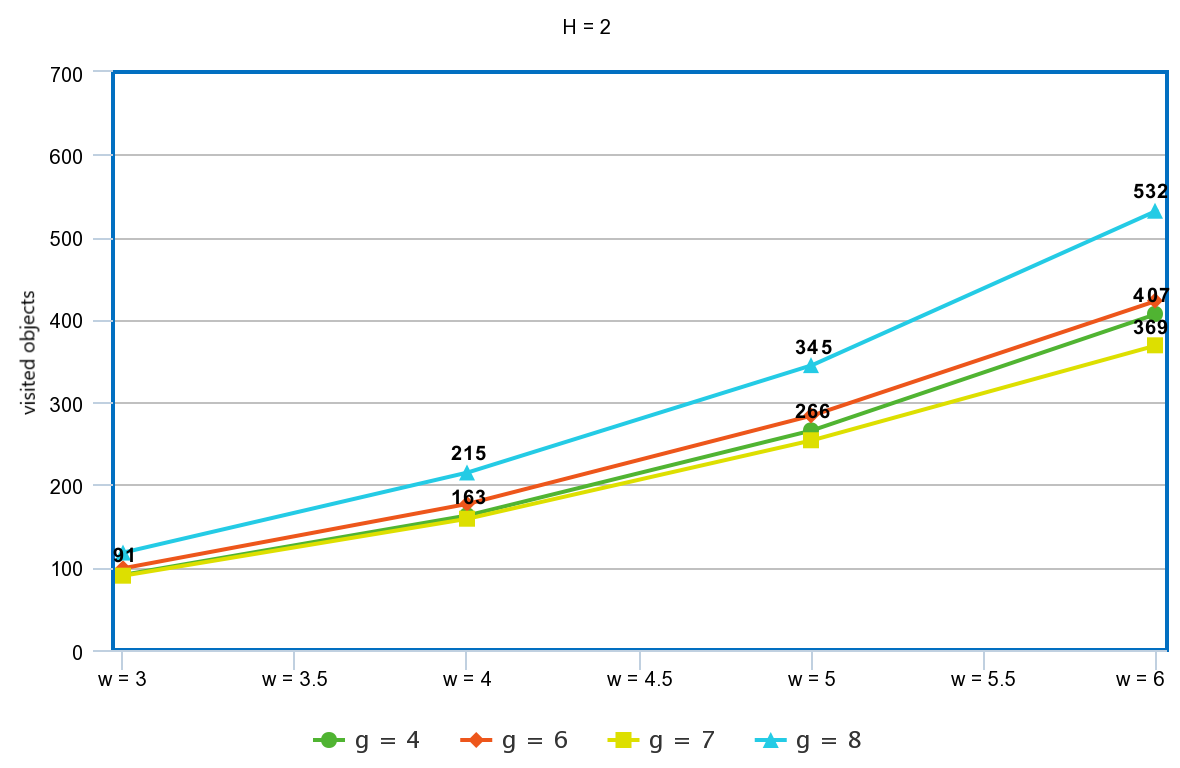}
        \caption{Resnet50v2 fine tuned model 2 - \textbf{visited objects} trend with fixed $h=2$}
    \end{minipage}
\end{figure}
\vspace{3mm}

\noindent The following figure shows the decrease of the mAP value when we increase the number of hyper-planes for each g-function. We can observe this trend, not only for the model in figure but also for all the other models we worked with.
\begin{figure}[H]
    \centering
    \begin{minipage}{0.5\textwidth}
        \includegraphics[width=1.\textwidth]{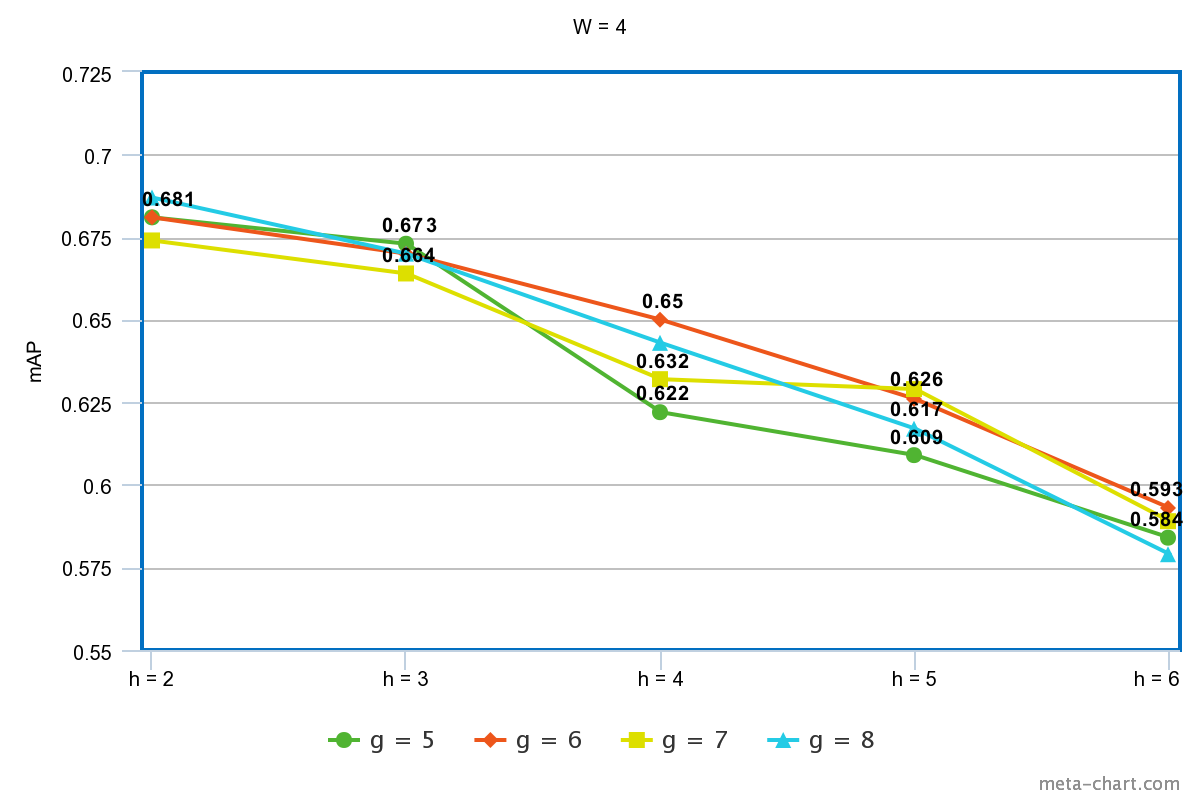}
    \end{minipage}
    \caption{Resnet50 feature extraction model - \textbf{mAP} trend with fixed $w=4$}
\vspace{2mm}
\end{figure}

\clearpage 

\section{Web application}\label{sec:7}
For the purposes of demonstrating our system, we developed a Web-App based on the best index created during the previous steps: the index was created using the 2nd model based on Resnet50v2 fine tuned as pre-trained neural network and the LSH index had the following hyper-parameters: \textbf{g=7}, \textbf{h=2}, \textbf{w=6}. For the demo we created a Python Flask project as web server, while the front-end was built using the framework \textit{Bootstrap}.

\begin{figure}[H]
    \centering
    \begin{minipage}{0.49\textwidth}
        \centering
        \includegraphics[width=1.\textwidth]{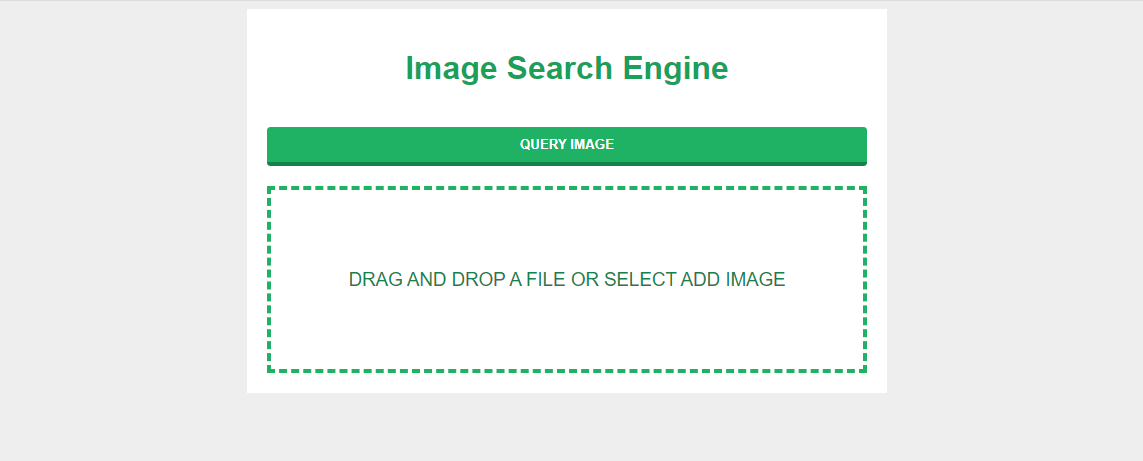}
        \caption{Web gui, home page}
    \end{minipage}\hfill
    \hspace{1mm}
    \begin{minipage}{0.49\textwidth}
        \centering
        \includegraphics[height=0.41\textwidth]{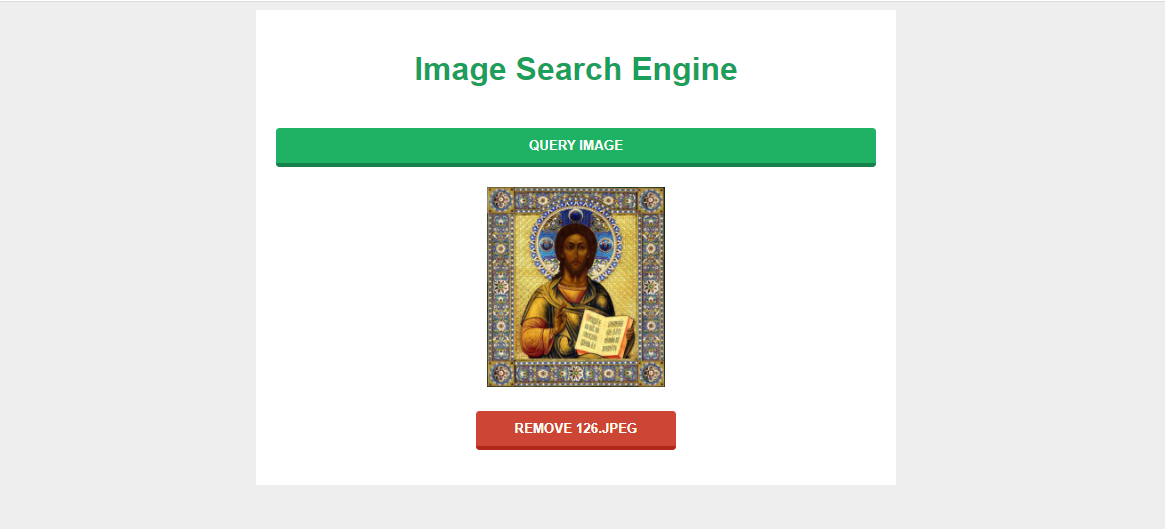}
        \caption{Web gui, select image as query}
    \end{minipage}
\end{figure}
\vspace{3mm}

\begin{figure}[H]
    \centering
    \begin{minipage}{0.65\textwidth}
        \includegraphics[width=1.1\textwidth]{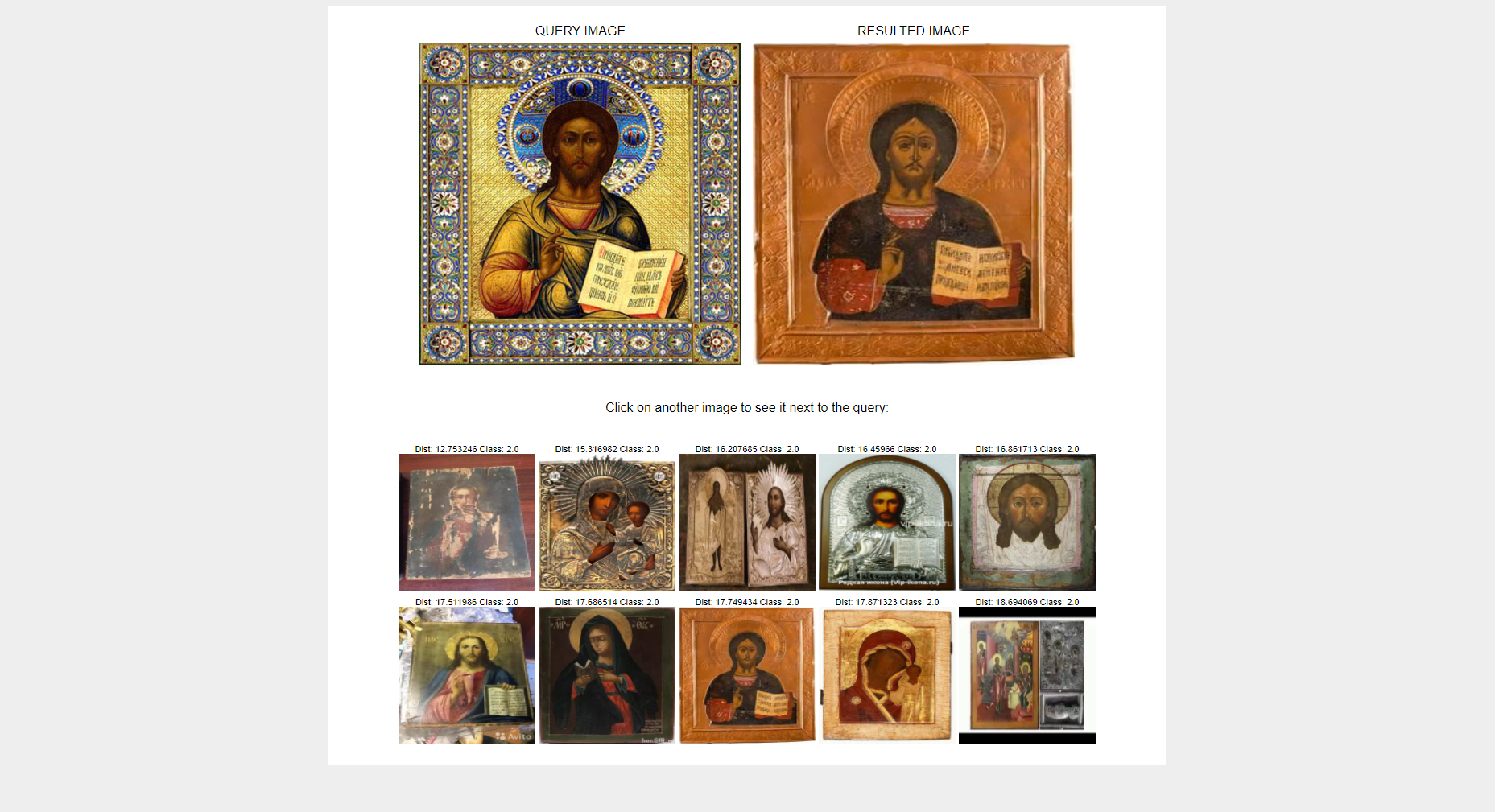}
    \end{minipage}
    \caption{Web gui, list of results}
\vspace{2mm}
\end{figure}

\section{Conclusions and Futuer Work}\label{sec:conclusion}
In this paper, we presented the design and development of web system for approximate image search using features extracted from an artificial neural network trained on artwork images taken from WikiArt archive \footnote{https://www.wikiart.org}. As an approximate index, we used our own implementation of the well-known Locality Sensitive Hashing. In the future, we may confront other approximate image search approaches, such as those exploiting surrogate texts \cite{amato2011combining,amato2016large,amato2020large,10.1145/3209978.3210089}.

\section*{Acknowledgement}
Thanks to professor Giuseppe Amato, professor Claudio Gennaro and professor Fabrizio Falchi for the course of "Multimedia Information Retrieval and Computer Vision" that gave us this opportunity.


\bibliographystyle{abbrv}
\bibliography{bibliography}

\end{document}